\documentclass[pra,aps,superscriptaddress]{revtex4-2}
\usepackage{graphicx}
\usepackage{epsfig}
\usepackage{graphics}
\usepackage{multirow}
\usepackage{dcolumn}
\usepackage{bm}
\usepackage{lipsum}
\usepackage{ragged2e}
\usepackage{braket}
\usepackage{bbm}
\usepackage{color}
\usepackage{xcolor}
\usepackage{amsfonts}
\usepackage{amsmath}
\usepackage{amsmath,amssymb,latexsym}
\usepackage{wasysym,stmaryrd,marvosym}
\usepackage{mathrsfs}
\usepackage{natbib}
\usepackage[thinlines]{easytable}

\usepackage[mathlines]{lineno}
\usepackage[colorlinks=true, allcolors=blue]{hyperref}
\newcommand{\comment}[1]{}
\usepackage{array}

\begin{document}
\title{Generalized rotating-wave approximation for the quantum Rabi model with optomechanical interaction}

\author{Wallace H. Monta\~no}
\affiliation{Centro de Nanociencias y Nanotecnolog\'ia,
Universidad Nacional Aut\'onoma de M\'exico, Apartado Postal 2681, 22800, Ensenada, Baja California, M\'exico.}

\author{Jes\'us A. Maytorena}
\affiliation{Centro de Nanociencias y Nanotecnolog\'ia,
Universidad Nacional Aut\'onoma de M\'exico, Apartado Postal 2681, 22800, Ensenada, Baja California, M\'exico.}

\date{\today}

\begin{abstract}
We investigate the spectrum of energy and eigenstates of a hybrid
cavity optomechanical system, where
a cavity field mode interacts with a mechanical mode of a vibrating 
end mirror via radiation pressure and with a two level atom via 
electric dipole interaction. 
In the spirit of approximations 
developed for the quantum Rabi model beyond rotating-wave approximation (RWA),
we implement the so-called generalized
RWA (GRWA) to diagonalize the tripartite Hamiltonian for arbitrary 
large couplings. Notably, the GRWA approach still allows to rewrite the 
hybrid Hamiltonian in a bipartite form, like a Rabi model with
dressed atom-field states (polaritons) coupled to mechanical modes through reparametrized 
coupling strenght and Rabi frequency.
We found a more accurate energy spectrum for a wide
range of values of the atom-photon and photon-phonon couplings,
when compared to the RWA results. The fidelity between 
the numerical eigenstates and its approximated counterparts is
also calculated. 
The degree of polariton-phonon entanglement of the eigenstates presents a non-monotonic behavior as the atom-photon coupling varies, in contrast to the characteristic monotonic increase
in the RWA treatment.
\end{abstract}

\maketitle

\section{INTRODUCTION}

Cavity quantum electrodynamics (cQED) and cavity optomechanics are two 
paradigmatic fields to study light-matter interaction and the coupling between
confined optical field and mechanical degrees of freedom, at a full quantum level of description.
The simplest model in cQED is the celebrated quantum Rabi model (QRM), which describes the interaction between a two-level quantum system (the ``atom'') and a single mode of a bosonic field \cite{PhysRev.49.324,PhysRev.51.652,Xie_2017}. It has been extensively
applied in a wide variety of fields, ranging from solid state \citep{PhysRevLett.110.066802,Yoshihara_2018} and quantum optics \citep{PhysRevLett.76.1800, PhysRevLett.68.1132,Trav_nec_2011,Crespi_2012} to molecular \cite{PhysRevA.41.1556,PhysRevA.65.063401,Albert_2012} and trapped ions physics \cite{RevModPhys.75.281, Pedernales2015,Cai_2021,Lv_2018} and recently, it has been proposed as basis for the development of quantum gates \cite{Schmidt_Kaler_2003,Moya_Cessa_2016}, information protocols \cite{Nielsen2012} and nondemolition measurement \cite{PhysRevLett.94.123602,Lindstr_m_2007}. The simplest counterpart in optomechanics is the model
consisting of an optical cavity with photons exerting radiation pressure 
through reflection on a movable end mirror. This model and related variants
have played a role in the goal 
of reaching quantum control of mechanical motion and detect small forces or displacements beyond the standard quantum limit \cite{Qoptomechanics}, and also
in a number of applications like gravitational wave astronomy \cite{2004JOptB...6S.675C,Abramovici1992}, cold atom experiments and creation of nonclassical macroscopic states \cite{Non-classical}. 

Recently these models have been combined in an analytically
solvable
atom-photon-oscillator model \cite{PhysRevLett.112.013601}. Its cQED part was described by the Jaynes-Cummings model (JCM), while the optomechanical part by a quantum harmonic oscillator whose canonical position  operator is coupled to the photon number operator. The model predicts
cooling of mechanical motion at the single-polariton level, antibunched states of mechanical motion, and quantum interference and correlation effects due to the
interplay of all the coupling mechanisms \cite{PhysRevA.95.023832}.  

As is well known, the JCM is derived from the QRM by using the rotating-wave approximation (RWA) \cite{Shore_1993,2012EPJST.203..163B}, which works well under quasi-resonance condition and for
small atom-field coupling strength compared to the energy of the photon field.
This approximation neglects the anti-resonant terms and restricts the
Hilbert space to an infinity set of 2D subspaces, each one characterized by the conservation of the
number of excitations \cite{Shore_1993,Braak_2019,JC_Journal}. Over the past decade,
the so-called ultrastrong and deep strong coupling regimes of the light-matter interaction have been reached in experiments with superconducting circuits,
semiconductor quantum wells, optomechanical systems, and other hybrid platforms \cite{RevModPhys.91.025005,Frisk_Kockum_2019}. On the theoretical side, this motivated the
development of several approximation methods for the QRM beyond the RWA
\cite{PhysRevB.72.195410,PhysRevLett.99.173601,PhysRevA.84.042110,disp1,
PhysRevLett.105.263603}. The exact solution of the QRM, for arbitrary magnitude
of the parameters, is already know since more than a decade \cite{Xie_2017,PhysRevLett.107.100401}; it is given in terms of the poles of transcendental functions, ant it lacks of analytical expressions of the spectrum of energy and eigenstates. Thus, it is still convenient to appeal to approximated models that lead to
 intuitive understanding of the physics, like in the RWA approach \cite{solano2011dialogue}.

Here, motivated by the work of Restrepo et al. \cite{PhysRevLett.112.013601},
we calculate the spectrum of energy, the eigenfunctions and its entanglement
properties, of the same atom-photon-oscillator model, but with the cQED part
described by the QRM and for arbitrary values of the atom-photon and
photon-phonon coupling strengths. To this end, we adapted the generalized
version of the RWA (GRWA), developed by K. Irish \cite{PhysRevLett.99.173601} for the QRM, to the full hybrid Hamiltonian. As a result, we found that the approach leads to accurate
expressions of the levels and states, in the whole range of couplings
and for large detuning, and to a degree of entanglement which differs
strongly from the RWA results.

The paper is organized as follows. In Section II we introduced the
hybrid Hamiltonian, and developed the GRWA strategy to obtain an
approximated Hamiltonian and its corresponding spectrum of energy
and states. In Section III a discussion of the entanglement properties
of the eigenstates is presented, based on the comparison between the
RWA and GRWA results for both, the QRM and the hybrid system.
Section IV is devoted to summarize our findings. Appendix A enumerates
some basis states associated to the GRWA for the QRM and needed for
the application of the GRWA approach to the hybrid model under consideration.

\section{Atom-photon-oscillator system with arbitrary large couplings}

\subsection{The model}

\begin{figure}[ht] 
\includegraphics[width=8cm]{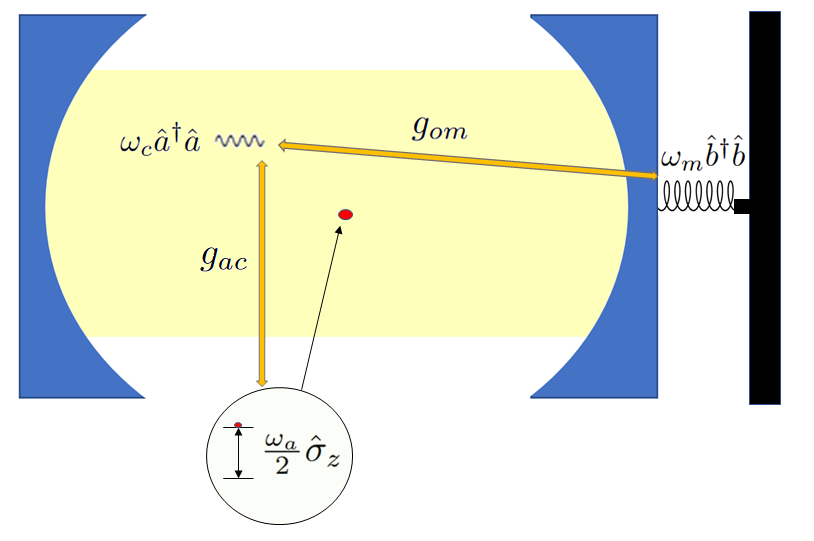}
\caption{Hybrid system combining the atom-photon interaction and
quantum optomechanics. A two-level system (the ``atom'') lies within a 
optical cavity (photons)
one of whose mirrors oscillates harmonically (phonons).} \label{Figura1}
\end{figure}

We consider a hybrid system that combines cQED and cavity optomechanics, with Hamiltonian (Fig.\,\ref{Figura1})
\begin{equation} \label{full_hyb}
    \hat{H}_{hyb} = \frac{\omega_a}{2} \hat{\sigma}_z+\omega_c \, \hat{a}^\dagger \hat{a} + \, g_{ac} \hat{\sigma}_x \, (\hat{a}^\dagger + \hat{a} )  + \omega_m \hat{b}^\dagger \hat{b} - g_{om} \hat{a}^\dagger \hat{a} \bigl(\hat{b}^\dagger + \hat{b} \bigl) .
\end{equation}
The first three terms corresponds to the well known quantum Rabi model\cite{PhysRev.49.324,PhysRev.51.652,Xie_2017} $\hat{H}_R$, 
where a two-level atom (level spacing $\omega_a$) is coupled (parameter $g_{ac}$) to a single-mode quantized electromagnetic field of an optical cavity (frequency $\omega_c$).
The fourth and fifth terms describe the standard optomechanical model, which assumes that one of the cavity mirrors oscillates harmonically
(frequency $\omega_m$) due to radiation pressure caused by the interchange of momentum of bouncing photons\cite{braginsky1967classical} (coupling $g_{om}$). The boson operators $\hat{a},\hat{a}^{\dag}$ ($\hat{b},\hat{b}^{\dag}$) are the annihilation and creation
operators for photons (phonons), and $\hat{\sigma}_i$ are the spin-$1/2$ Pauli matrices; we use $\hbar=1$. For the sake of simplicity, the atom-oscillator coupling 
is not included in model (\ref{full_hyb}), however it can be added without major changes in the derivation that follows.
        
The version of the hybrid system introduced by Restrepo et al. \cite{PhysRevLett.112.013601} 
used the JCM for the atom-photon part, instead of the QRM. 
The Jaynes-Cummings Hamiltonian 
$\hat{H}_{JC}=\frac{\omega_a}{2}\hat{\sigma}_z+\omega_c \hat{a}^\dagger \hat{a}+
g_{ac}(\hat{\sigma}_+\hat{a}+\hat{\sigma}_-\hat{a}^{\dag})$, where 
$\hat{\sigma}_{\pm}=(\hat{\sigma}_x\pm i\hat{\sigma}_y)/2$, neglects
the counter-rotating terms, and is valid for
weak coupling, $g_{ac}/\omega_c\ll 1$, and near resonance $\omega_c\approx\omega_a$ \cite{1443594}. In order to develop a generalized RWA for the hybrid system, its Hamiltonian must include the complete atom-photon interaction $\hat{H}_R$.

\subsection{GRWA Hamiltonian}

The use of the JCM in the cQED part of the Hamiltonian (\ref{full_hyb}) allows to break the
Hilbert space into a set of invariant subspaces characterized by the
conservation of the atom-cavity excitation number $\hat{N}_{JC} = \hat{a}^\dagger \hat{a} + (\hat{\sigma}_z+1)/2$   \cite{PhysRevLett.112.013601,PhysRevA.95.023832}. When $\hat{N}_{JC}\neq 0$, further order-of-magnitude simplifications lead to a mathematical structure in which each subspace writes as a Rabi Hamiltonian. At this point, an additional RWA to neglect counter-rotating terms in the effective bipartite system formed by the atom-cavity dressed states (polaritons) and the mechanical vibrations, implies further reduction to $2\times 2$ blocks of the JC type \cite{PhysRevLett.112.013601,PhysRevA.95.023832}. This allows in turn to obtain analytical expressions for the non-trivial part of the energy spectrum \cite{PhysRevLett.112.013601,PhysRevA.95.023832}.
The case for $\hat{N}_{JC}=0$ corresponds to a subspace of a 
simple displaced harmonic oscillator in the phonon operators.

In what follows, we adopt a similar procedure to derive an approximate Hamiltonian
to model (\ref{full_hyb}), but introducing a generalized version of the RWA (the GRWA)
instead to proceed within the weak coupling and quasi-resonant restrictions.
The GRWA is an approach to the QRM developed to go beyond the JCM conditions, allowing to
explore a wider range of frequency detuning values and larger magnitudes of the
atom-cavity coupling \cite{PhysRevLett.99.173601}. As we will see, within the context of the hybrid model (\ref{full_hyb}), this approximation allows also a better description of the
atom-cavity polaritons, the generalized Rabi frequency,
energy spectrum and eigenstates, than its RWA counterpart.

As will be discussed below, the conserved quantity within the GRWA is the excitation
number involving adiabatic states and displaced photons. Correspondingly, the atom-cavity polaritonic states becomes dressed differently with respect to the RWA. On the other hand, the optomechanical interaction mixes the GRWA polaritonic states of different subspaces. In order to obtain disconnected invariant subspaces,
the GRWA should be applied to the complete
Hamiltonian (\ref{full_hyb}). In our derivation we will use
several quantities associated to the GRWA of the QRM, like the displaced oscillator
basis or the adiabatic basis states, the corresponding energies, and frequency parameters. These are presented in an appendix, to simplify the application
of the GRWA to the full Hamiltonian (\ref{full_hyb}).

We start by writing an approximate (GRWA) Hamiltonian for the Rabi part
of the tripartite model (\ref{full_hyb}), through the spectral decomposition
\begin{equation} \label{HGRWA}
    \hat{H}_R^{grwa} = E^{grwa}_G \hat{P}_G + \sum^{\infty}_{N=0} \left[ E_{+,N}^{grwa} \hat{P}_{+,N} + E_{-,N}^{grwa} \hat{P}_{-,N} \right],
\end{equation}
which involves the energy of the ground state $E^{grwa}_G$, the doublet 
$E_{\pm,N}^{grwa}$, and the projectors $\hat{P}_G = \ket{\psi^{grwa}_{G}}\bra{\psi^{grwa}_{G}}$, $\,\hat{P}_{\pm,N} = \ket{\psi^{grwa}_{\pm,N}}\bra{\psi^{grwa}_{\pm,N}}$, in terms of the GRWA basis 
\cite{PhysRevLett.99.173601} $\{|\psi_G^{grwa}\rangle, \ket{\psi^{grwa}_{\pm,N}}, N=0,1,\ldots\}$
(see Appendix A). 

The hybrid Hamiltonian (\ref{full_hyb}) is now written as
$\hat{H}_{hyb} \approx \hat{H}^{grwa}_{hyb}$, where
\begin{equation}
\hat{H}^{grwa}_{hyb} = \hat{H}_R^{grwa} + \hat{H}_{om},
\label{333}
\end{equation}
with the last term containing the non-Rabi contribution, namely
the optomechanical part $\hat{H}_{om}=\omega_m \hat{b}^\dagger \hat{b} - g_{om} \hat{a}^\dagger \hat{a} (\hat{b}^\dagger + \hat{b})$. The substitution $\hat{H}_R^{grwa}\to\hat{H}_{JC}$
should produce the results reported in references \onlinecite{PhysRevLett.112.013601,PhysRevA.95.023832}.
As is well known, the GRWA breaks the Rabi Hamiltonian $\hat{H}_R$ into $2\times 2$
invariant subspaces (see Appendix A), and we note that each one is characterized by a conserved number
given by  
\begin{equation} \label{Nnew}
 \hat{N}_R^{grwa} = \frac{1}{2} \bigl( \hat{I}^{(2)} + \hat{\sigma}_x  \bigl) D^\dagger(g_{ac}/\omega_c) \hat{a}^\dagger \hat{a}  D(g_{ac}/\omega_c) + \frac{1}{2}\bigl(\hat{I}^{(2)} - \hat{\sigma}_x  \bigl) D(g_{ac}/\omega_c) \hat{a}^\dagger \hat{a}  D^\dagger(g_{ac}/\omega_c) + \sum_N \ket{\psi^{ad}_{+,N}}\bra{\psi^{ad}_{+,N}},     
\end{equation}
where $D(\pm \nu)= e^{\pm \nu(\hat{a}^\dagger - \hat{a})}$ is the displacement operator and $\hat{I}^{(2)}$ is the $2\times 2$ identity matrix. We recall that the starting point
of the GRWA is to express the Rabi Hamiltonian $\hat{H}_R$ in the adiabatic
basis \cite{PhysRevB.72.195410} $\{\ket{\psi^{ad}_{\pm,N}}, N=0,1,\ldots\}$ (see Appendix A), instead of using the eigenbasis of the noninteracting Hamiltonian
$\frac{\omega_a}{2} \hat{\sigma}_z+\omega_c\hat{a}^\dagger \hat{a}$. 
As a consequence, the number (\ref{Nnew}) involves $\hat{\sigma}_x$ and the adiabatic states $\ket{\psi^{ad}_{+,N}}$. 
Note however that $[\hat{H}_{om},\hat{N}_R^{grwa}]\neq 0$, and then the optomechanical part, after introducing the projectors used in (\ref{HGRWA}), becomes
\begin{equation}
\begin{split}
\hat{H}_{om}  =  \omega_m  \hat{b}^{\dagger} \hat{b}  - (\hat{b}^{\dagger} + \hat{b})\left[ \left(\hat{P}_G + \sum^{\infty}_{N=0}(\hat{P}_{+,N} + \hat{P}_{-,N})\right) g_{om} \, \hat{a}^{\dagger} \hat{a}  \left(\hat{P}_G + \sum^{\infty}_{N'=0}(\hat{P}_{+,N'} + \hat{P}_{-,N'} )\right) \right].
\label{non-rabi}
\end{split}
\end{equation}
 These terms connect projectors with different $N$ index, having terms like $ g_{om} (\hat{b}^{\dagger} + \hat{b}) \hat{P}_{-,N} (\hat{a}^\dagger  \hat{a})  \hat{P}_{-,N+2} \neq 0$ or 
$ g_{om} (\hat{b}^{\dagger} + \hat{b}) \hat{P}_{-,N} (\hat{a}^\dagger  \hat{a})  \hat{P}_{+,N+2} \neq 0$ 
 , among others. Proceeding in the same spirit of the GRWA \cite{PhysRevLett.99.173601}, these crossed terms involving different labels $N$ will be neglected. 
 As we will show below, this help us to find and approximated Hamiltonian which
 commutes with the operator $\hat{N}^{grwa}_R$ (\ref{Nnew}), in complete analogy
 to the hybrid RWA Hamiltonian of Ref.\onlinecite{PhysRevLett.112.013601} which commutes with the conserved number $\hat{N}_{JC}$ of the JCM. Such a simplification of (\ref{non-rabi})
allows to breaks the Hilbert space into a set
of disconnected invariant subspaces, each with 
$N_R^{grwa}\equiv\langle\psi^{grwa}_{\pm,N}|\hat{N}_R^{grwa}|
\psi^{grwa}_{\pm,N}\rangle=N+1$ 
atom-cavity polaritons. 
Therefore, the Hamiltonian $\hat{H}^{grwa}_{hyb}$ will have projectors involving only the
same index $N$, with the state $\ket{\psi^{grwa}_{\pm,N}}$
 containing exactly $N_R^{grwa}$ polaritons, while the ground state  
$\ket{\psi^{grwa}_{G}}$ contains zero.
Droping the mixing terms is equivalent to perform the GRWA to $\hat{H}_{om}$. This contrasts to the RWA approach \cite{PhysRevLett.112.013601,PhysRevA.95.023832}, where such a mixing is absent.

The photon number operator in the optomechanical term 
become
\begin{equation}
\hat{a}^{\dagger} \hat{a} =  \frac{g_{ac}^2}{\omega_c^2}  \hat{P}_G + \sum^{\infty}_{N=0} \left[  \left( N + 1/2 + \frac{g_{ac}^2} {\omega_c^2} \right) \hat{I}^{(N)} + \frac{1}{2} \left(\cos{\alpha_N}  - \frac{\Omega_N }{\omega_c} \sin{\alpha_N}  \right)\hat{\sigma}^{(N)}_z
-  \frac{1}{2} \left(\sin{\alpha_N} + \frac{\Omega_N}{\omega_c} \cos{\alpha_N} \right)  \hat{\sigma}^{(N)}_x \right]
\end{equation}
where $\Omega_N= 2 g_{ac}\sqrt{N+1}$ is the characteristic Rabi frequency for
$N$ photons of the JCM \cite{2012EPJST.203..163B}. The angle 
$\alpha_N$ is defined by $\tan{\alpha_N} = -\Omega_{N,N+1}/\Delta_N$, 
with $\Omega_{N,N'}= \omega_a \braket{N_-|N'_+}$ and
$\Delta_N =  \frac{1}{2}\left[\Omega_{N,N}+\Omega_{N+1,N+1} \right] - \, \omega_c $, where $\langle N_{-}|N^\prime_{+}\rangle$
is the overlap between oppositely displaced Fock states
(see (\ref{overlapMN}) and states (\ref{N+-}) in Appendix A). To write these expressions, we defined Pauli matrices $\hat{\sigma}^{(N)}_{x,y,z}$ from the doublet $\ket{\psi^{grwa}_{\pm,N}}$, and the identity $\hat{I}^{(N)} = \hat{P}_{+,N} + \hat{P}_{-,N} $.

In order to rationalize what is neglected it is convenient
to move to the interaction picture through the unitary transformation 
$\hat{U}(t)=\exp{(i \hat{H}_R^{grwa}t)}$. After this, each term of the effective 
Hamiltonian $\hat{U}(t) \hat{H}_{om} \hat{U}^\dagger(t)$  rotates at a speed determined by the energy difference of the GRWA states that it connects. 
For a given $N$, the diagonal terms become time independent and off-diagonal terms oscillating with the generalized GRWA Rabi frequency 
$T_N=\sqrt{\Delta_N^2+\Omega_{N,N+1}^2}$ (Eq.(\ref{GTN})). 
In addition, there also appear non diagonal
terms connecting states with different quantum numbers $N\neq N'$, 
revolving at higher frequencies $\omega_c (N-N^\prime ) + (T_N \pm T_{N^\prime})/2$.
These terms are non-energy-conserving and will be neglected in the same way as the counter-rotating terms arising in the GRWA for the Rabi Hamiltonian $\hat{H}_R$.

As a consequence, the hybrid system can be viewed as two coupled subsystems defined
by the atom-cavity polaritons (a dressed two-level system for a given $N$ in the GRWA basis) and by displaced phonons. 
The Hilbert space reads as a direct sum of invariant subspaces
$\mathcal{H} = \mathcal{H}_G \oplus \sum^{\infty}_{N=0} \mathcal{H}^{(N)}$.
Correspondingly, the total Hamiltonian, as obtained from (\ref{HGRWA}) and (\ref{333})
without mixing terms, reads as
\begin{equation} \label{totalhyb33}
\hat{H}^{grwa}_{hyb} = \hat{H}_G  + \sum^{\infty}_{N=0} \hat{H}^{(N)}\,,
\end{equation}
where 
\begin{equation} \label{basebase2}
\hat{H}_G= \left[\ E^{grwa}_{ G}+ \omega_m \hat{b}^{\dagger} \hat{b} -  g_{om}\frac{g_{ac}^2}{\omega_c^2} ( \hat{b}^{\dagger}+\hat{b})  \right] \hat{P}_{G},
\end{equation}
is associated to the polaritonic ground state, and
\begin{eqnarray}
\hat{H}^{(N)} & = & \Bigl[ \omega_m \hat{b}^{\dagger} \hat{b} +  k_N - q_N ( \hat{b}^\dagger +\hat{b}) \Bigl] \hat{I}^{(N)} + \left[\frac{T_N}{2} +  \frac{g_{om}}{2} \left(  \frac{\Omega_N}{\omega_c} \sin \alpha_N  - \cos \alpha_N \right)  (b^{\dagger} + b) \right] \hat{\sigma}^{(N)}_z 
\label{bloqueasd} \\
&& \hspace*{8cm} +  \frac{g_{om}}{2} \left( \sin \alpha_N +  \frac{\Omega_N}{\omega_c} \cos \alpha_N  \right) (b^{\dagger} + b) \hat{\sigma}_x^{(N)}, \nonumber
\end{eqnarray}
to an effective spin-boson Hamiltonian. 
The additive constant $k_N$ and the static shift of the mechanical resonator
$q_N$, for the subspace $N$, are given by
\begin{eqnarray}
k_N &=& \omega_c \left(N+ \frac{1}{2}  - \frac{g_{ac}^2}{\omega_c^2} \right)
+ \frac{1}{4}\left(\Omega_{N,N} -\Omega_{N+1,N+1}\right), \\
q_N &=&   g_{om} \left(N+\frac{1}{2} + \frac{g_{ac}^2}{\omega_c^2}\right)  
\label{qn}.
\end{eqnarray}

Note that the operator (\ref{Nnew}) remains a conserved quantity for the
hybrid GRWA Hamiltonian (\ref{totalhyb33}),
$[\hat{H}^{grwa}_{hyb},\hat{N}_R^{grwa}]=0$.
The first term on the RHS of (\ref{totalhyb33}) is just a displaced harmonic oscillator tensor product with the polaritonic ground state projector. 
Although the second term looks more complicated, note however that for each $N$
we have an interacting spin-boson like Hamiltonian that can be independently treated. 
To this end, we simplify the Hamiltonian (\ref{bloqueasd}) to a Rabi model-like form.
Note at this point that when $g_{ac}/\omega_c\ll 1$, all the quantities in our Hamiltonian (\ref{bloqueasd}) reduces to that reported in reference \onlinecite{PhysRevLett.112.013601}. 

\begin{figure}[ht]
\centering
\includegraphics[width=8.9cm]{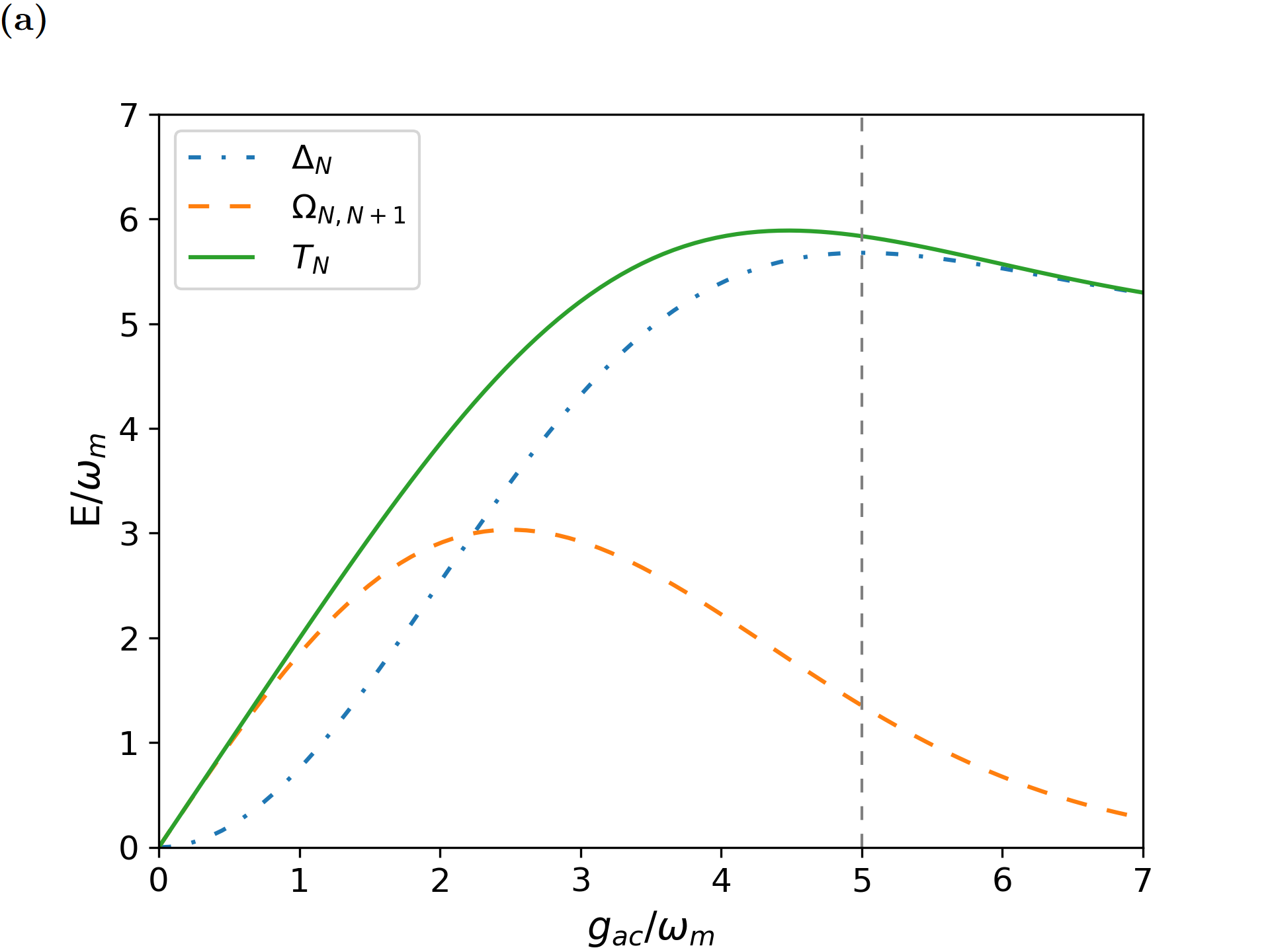}\includegraphics[width=8.9cm]{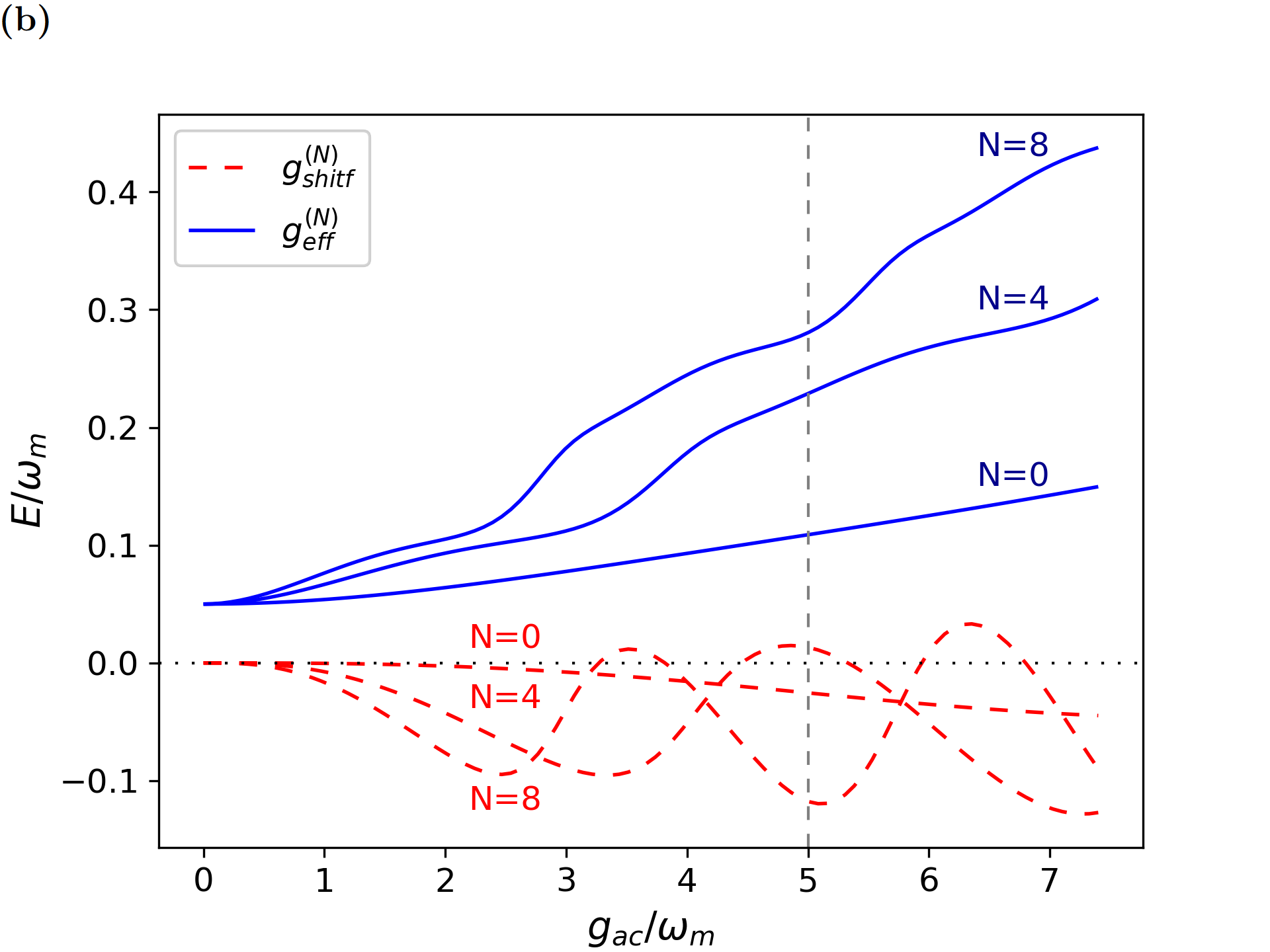}
\caption{Relevant GRWA parameters versus atom-photon coupling.
(a) Generalized GRWA Rabi frequency $T_N=\sqrt{\Delta_N^2+\Omega^2_{N,N+1}}$, 
for $\omega_c = \omega_a = 5 \omega_m$ and $N=0$. (b) Effective couplings $g^{(N)}_{shift}$ and $g^{(N)}_{eff}$ in the GRWA Hamiltonian $H^{(N)}$ (\ref{bloqueasd}). The vertical dashed line indicates the boundary between 
the ultra-strong and deep-strong coupling regimes, $g_{ac}=\omega_c$.}
\label{couplings}
\end{figure}

In contrast to the RWA version of the Eq. (\ref{bloqueasd}), where the quasi-resonance condition $|\omega_a - \omega_c| \ll \Omega_N$ allows to simplify the angle $\alpha_N$ to $\pi/2$ for all values of $g_{ac}$, the GRWA Hamiltonian involves an angle which depends on $\Delta_N$ and $\Omega_{N,N+1}$, and therefore such a simplification does not occur. 
Further reduction is achieved noting that, in the range of values of couplings and quantum number $N$ considered in this study,
the coupling factor
$g^{(N)}_{Shift} \equiv   \frac{g_{om}}{2} \bigl(  \frac{\Omega_N}{\omega_c} \sin \alpha_N  - \cos \alpha_N \bigl)$
is about an order of magnitude less than the generalized GRWA Rabi frequency $T_N$,
leaving the $\sigma_z^{(N)}$ term just as an energy shift (see Fig.\,\ref{couplings}(a)).
Note however that the effective coupling $g^{(N)}_{eff} \equiv \frac{g_{om}}{2} \bigl( \sin \alpha_N +  \frac{\Omega_N}{\omega_c} \cos \alpha_N  \bigl)$ should be kept for the whole
coupling range. Figure \ref{couplings}(b) show these quantities for $N=0,4,8$, at resonance $\omega_c=\omega_a$.

A convenient change to a displaced frame leads to the Rabi-like Hamiltonian
\begin{equation}
\hat{H}^{(N)}  =  C_N + \omega_m  \hat{b}_N^{\dagger} \hat{b}_N   + \frac{T_{N}}{2}  \hat{\sigma}_z^{(N)} +  {g}_{eff}^{(N)}  \Bigl( \hat{b}_N^{\dagger} + \hat{b}_N \Bigl) \hat{\sigma}_x^{(N)},
\label{hibi}
\end{equation}
where $\hat{b}_N = \hat{D}(q_N/\omega_m)\, \hat{b} \, \hat{D}^\dagger(q_N/\omega_m) = \hat{b} - q_N/\omega_m$. The first term is the constant $C_N = k_N - q_N^2 /\omega_m$. Following reference \onlinecite{PhysRevA.95.023832}, we have ignored in 
(\ref{hibi}) an off-diagonal term $(2q_N/\omega_m)g^{(N)}_{eff}\hat{\sigma}_x^{(N)}$,
which produces a Stark shift of the energies, in order to maintain analytical solvability.

Equations (\ref{basebase2}) and (\ref{hibi}) constitute the GRWA version of the
hybrid Hamiltonian (\ref{full_hyb}). In the following we use them
to calculate the energy spectrum and eigenstates for arbitrary large values of the
couplings.

\subsection{Eigenstates and energy spectrum}

We first consider the spectrum associated to the contribution
$\hat{H}_G$ (\ref{basebase2}), which involves the polaritonic ground state
term $\hat{P}_G$ and a displaced quantum harmonic oscillator. Thus, 
the corresponding energy eigenvalue is given by $E_G^{grwa}$ (see Eq.(\ref{g_egrwa})) plus the quantized energy spectrum of such an oscillator.
Explicitly, 
\begin{equation}
E_M = \omega_m  M - \frac{1}{\omega_m}\Bigl(\frac{g_{om} g_{ac}^2}{\omega_c^2}\Bigl)^2  
+\,E_G^{grwa},\ \ \  M=0,1,\ldots
\label{finalM}
\end{equation}
The associated eigenstates are
\begin{equation}
 \ket{\Psi_{M}} = \ket{\psi^{grwa}_{G}} \ket{M^{(q_{G})}},
\label{finalgrwa}
\end{equation}
where $\ket{\psi^{grwa}_{G}}$ is the ground state of the Rabi Hamiltonian under the GRWA (\ref{HGRWA}), and $\ket{M^{(q_{G})}}= D \left(q_{G}/\omega_m \right) \ket{M} $ is a phonon number state displaced by the quantity $q_{G}/\omega_m$, where
$q_{G}=  g_{om} g_{ac}^2/\omega_c^2 $.
We note that the state $\ket{\psi^{grwa}_{G}}$ coincide with the ground state
 of the Rabi Hamiltonian under the adiabatic approximation (see Eq.\,(\ref{g_sgrwa}))
 and contains exactly zero atom-cavity polaritons ($N_R^{grwa}=0$).

\begin{figure}
\centering
\includegraphics[width=8.8cm]{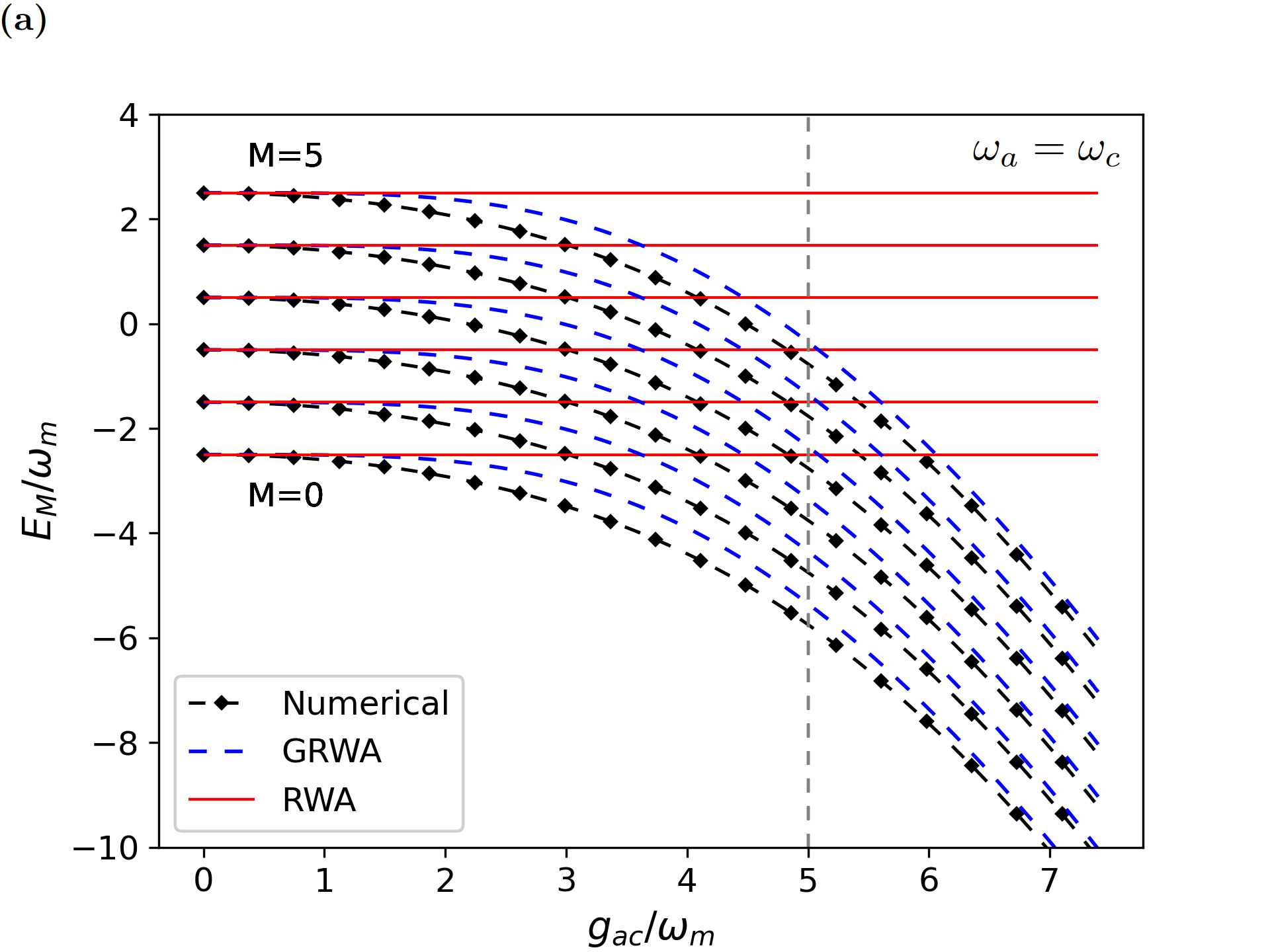}\includegraphics[width=8.8cm]{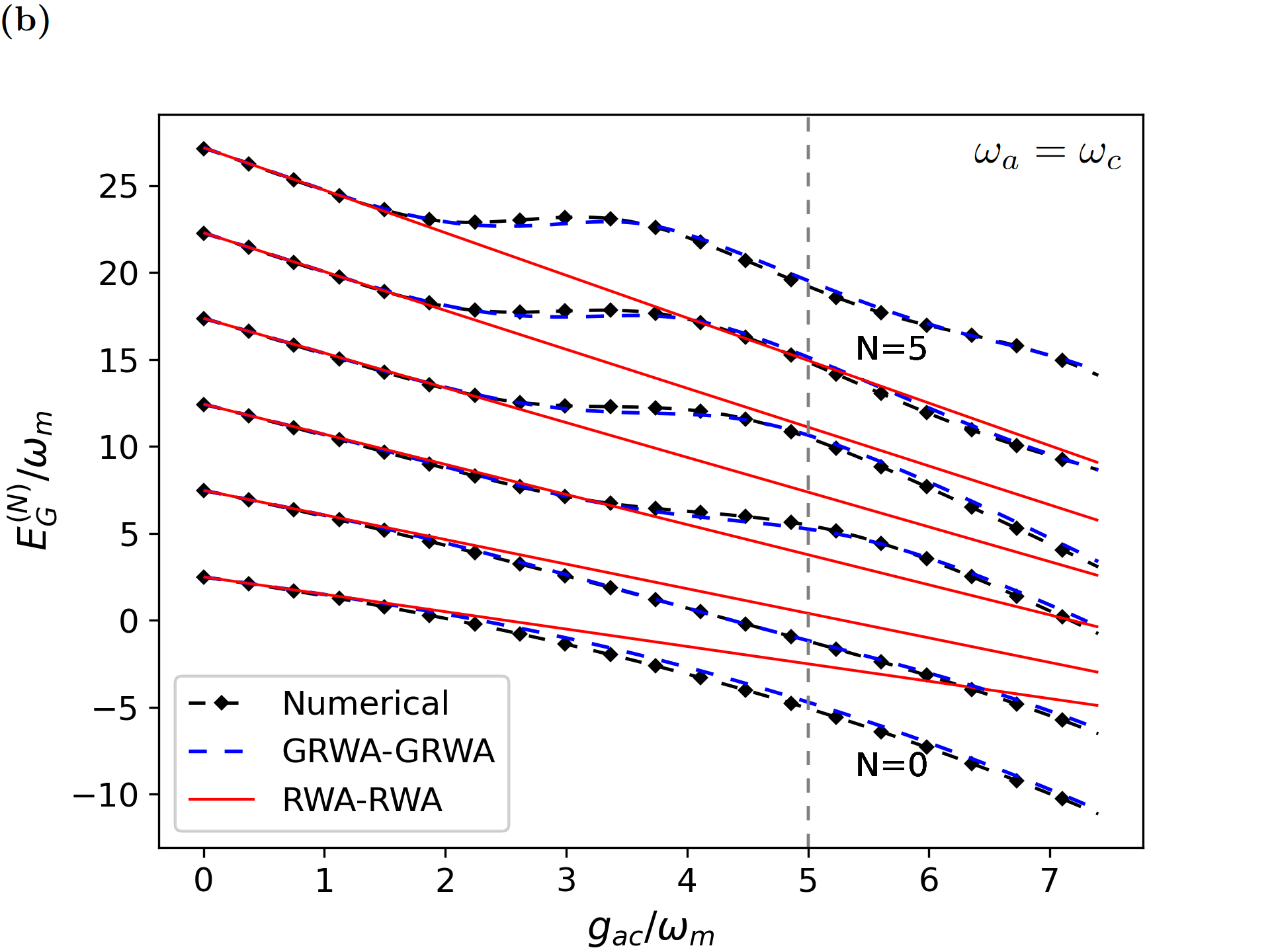}
\includegraphics[width=8.8cm]{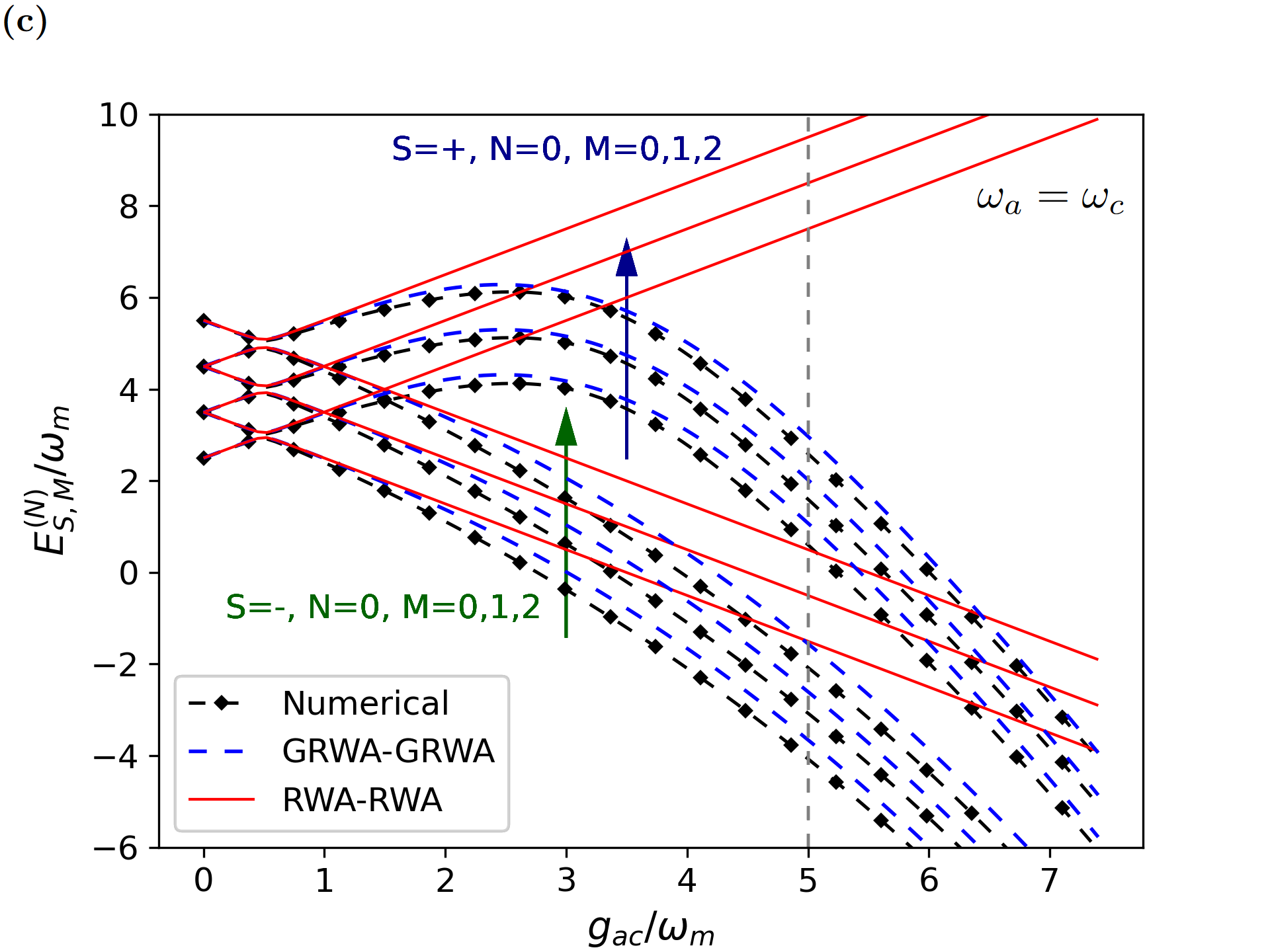}\includegraphics[width=8.8cm]{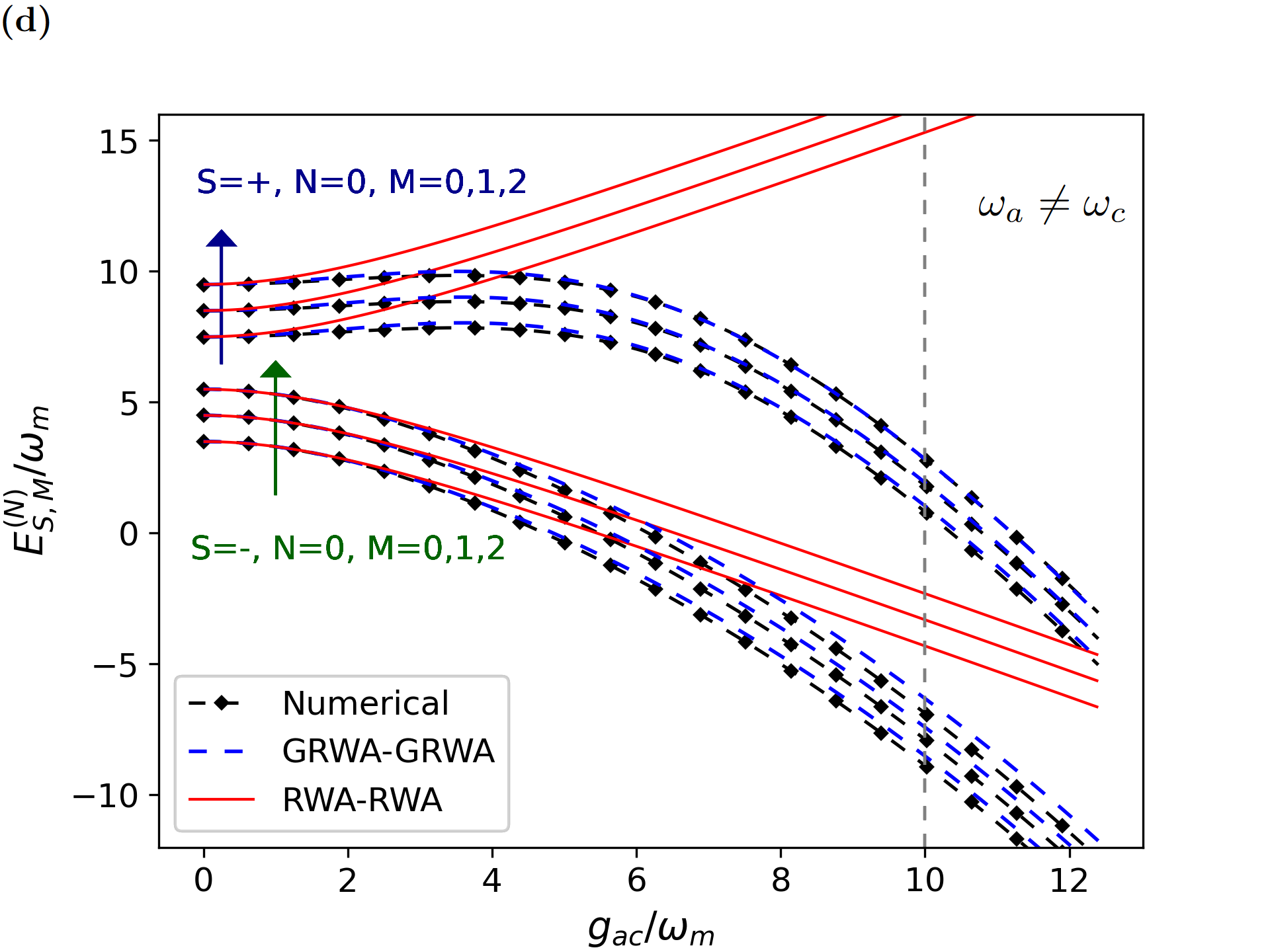}
\includegraphics[width=8.8cm]{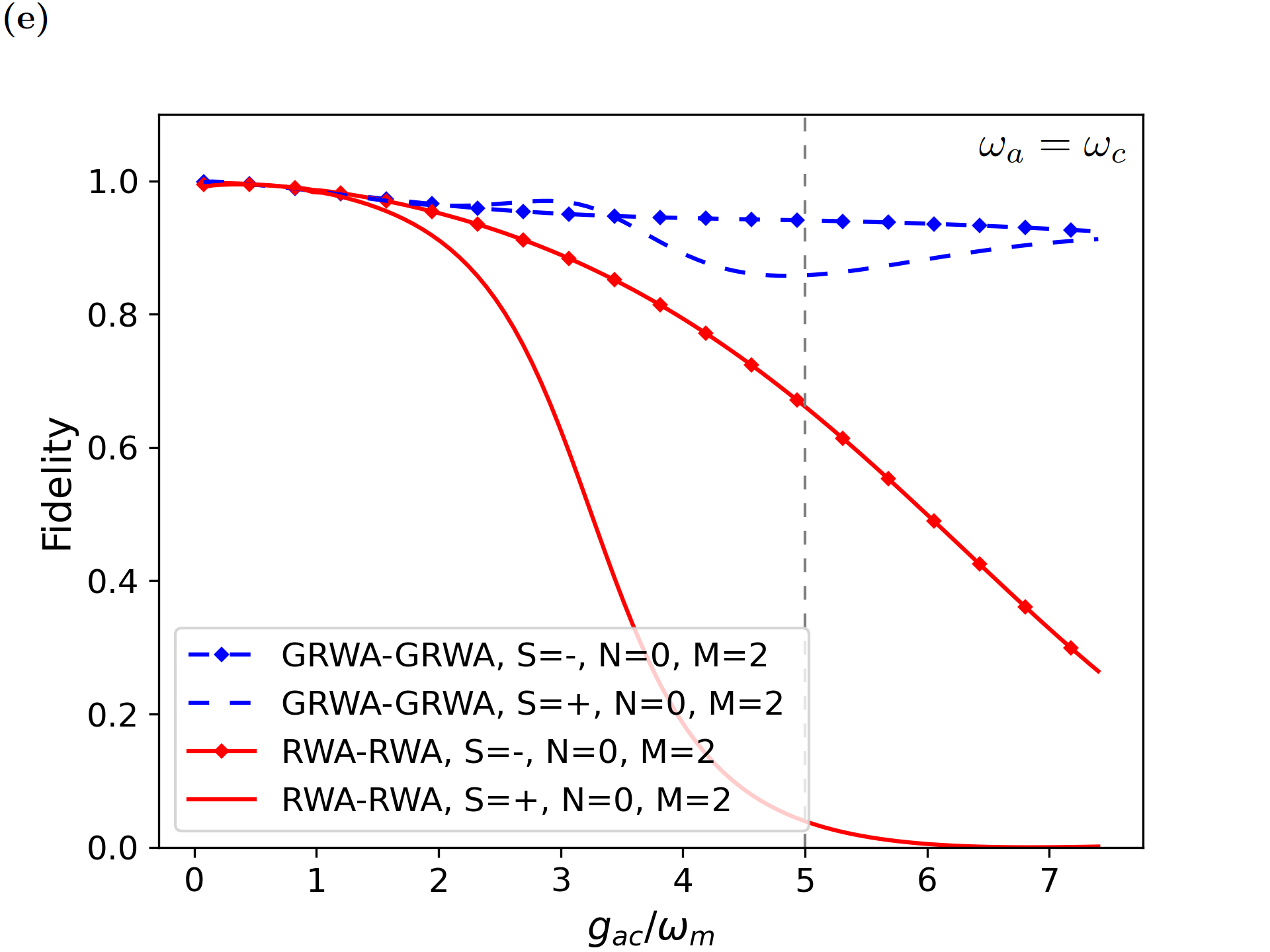}\includegraphics[width=8.8cm]{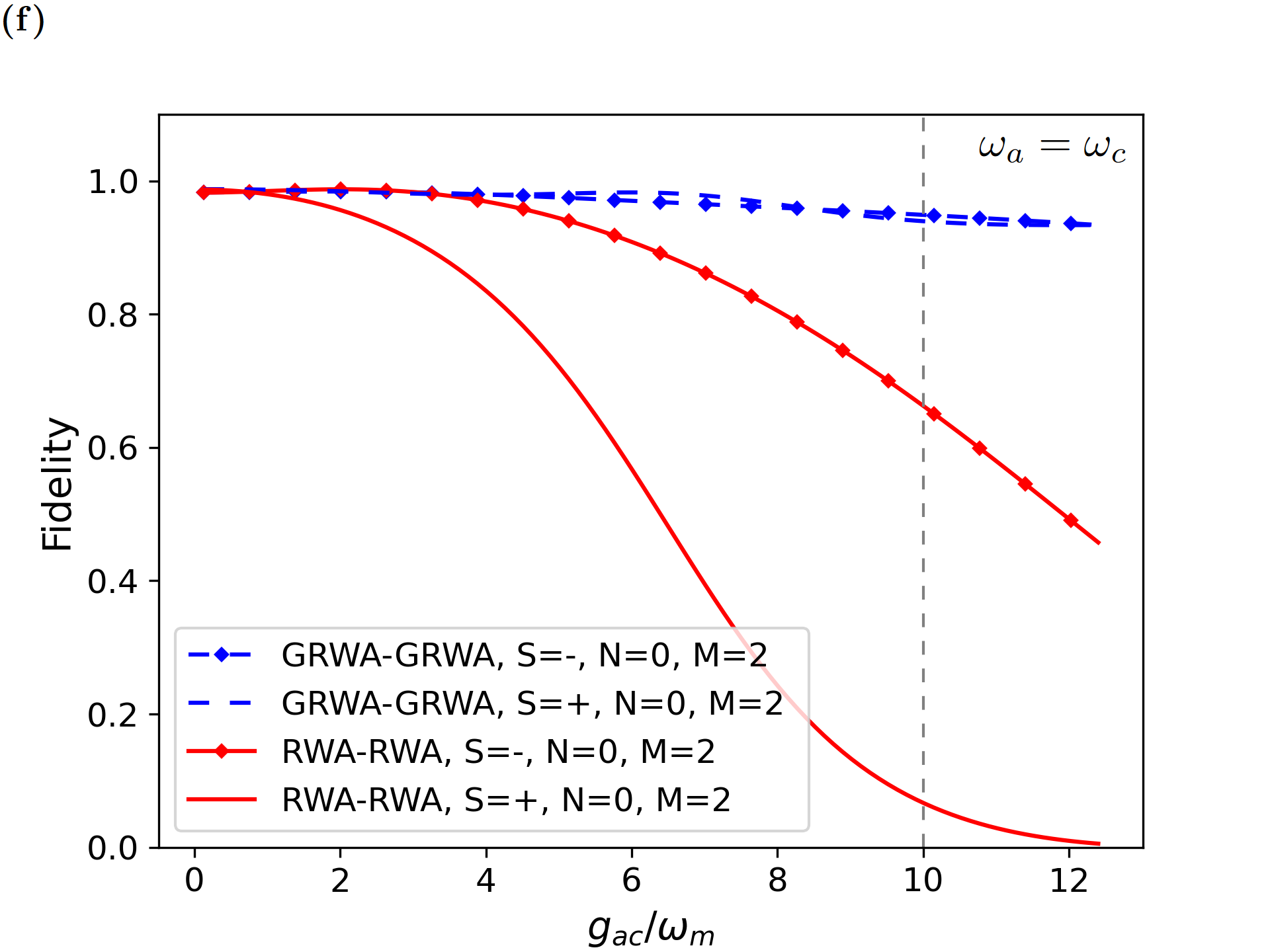}
\caption{Low energy levels versus atom-cavity $g_{ac}$ of the hybrid Hamiltonian within
the RWA-RWA (red solid line) and the GRWA-GRWA (blue dashed line) approaches, 
and the numerical exact solution (black dotted-dashed line).
(a) The zero atom-cavity polaritons energy $E_M$ (\ref{finalM}) for $M=0,1,\ldots 5$. (b) The zero phonons energy $E_G^{(N)}$ (\ref{grwadim1}) for $N=0,1,\ldots 5$. (c) The doublet $E^{(N)}_{\pm,M}$
(\ref{iniciogrwa}). In (a),(b) and (c) results at resonance
$\omega_c=\omega_a= 5 \, \omega_m $ are displayed, while (d) presents
the off-resonant case with $\omega_c = 5 \, \omega_m,\,\omega_a= 10 \, \omega_m$. In all cases we take $g_{om}=0.1\omega_m$. The fidelity
$| \braket{\Psi^{(0)}_{\pm,2} |\Psi^{numerical}}|^2$ of the states in (c) and (d)
are shown in (e) and (f), respectively.  The vertical dashed line indicates the boundary between 
the ultra-strong and deep-strong coupling regimes, $g_{ac}=\omega_c$.}
\label{plot_energias}
\end{figure}

Figure \ref{plot_energias}(a) shows the GRWA energy (\ref{finalM}) for
several values of the phonon quantum number $M$ as a function of
the atom-cavity coupling, the corresponding
RWA spectra \cite{PhysRevA.95.023832}, and the values obtained from a
numerical evaluation of the spectrum of Hamiltonian (\ref{full_hyb}),
at resonance $\omega_c = \omega_a = 5\, \omega_m$ 
with $g_{om}=0.1 \omega_m$. For small $g_{ac}$, the 
RWA and GRWA energies coincide and both fit well the numerical results.
At intermediate values of $g_{ac}$ the RWA lines remain constant while the
GRWA curves follow the exact (numerical) results, with the agreement
improving for larger couplings and for $\omega_c\neq\omega_a$.   

The remaining part of the spectrum corresponds to the states of the 
Rabi-like polariton-phonon Hamiltonian $\hat{H}^{(N)}$ (\ref{hibi}). 
For each $N$, there will be an isolated state and a set of doublets with 
dressed polariton-phonon states,
in analogy to the well known Jayne-Cummings or the GRWA spectrum \cite{PhysRevLett.99.173601}.
The former is given by
\begin{equation}
\ket{\Psi_G^{(N)}} = \frac{1}{\sqrt{2}} \bigl[ \ket{\psi^{grwa (x)}_{+,N}} \ket{(M=0)^{(q_N)}_{+^\prime}}  - \ket{\psi^{grwa (x)}_{-,N}}   \ket{(M=0)^{(q_N)}_{-^\prime} } \bigl],
\label{2323232}
\end{equation}
where 
\begin{equation}
\ket{\psi^{grwa (x)}_{\pm,N}} =  \frac{1}{\sqrt{2}} \bigl[ \ket{\psi^{grwa}_{+,N}} \pm \ket{\psi^{grwa}_{-,N}} \bigl]
\end{equation}
are the eigenkets
of $\hat{\sigma_x}^{(N)}=|\psi^{grwa}_{+,N}\rangle\langle\psi^{grwa}_{-,N}|+
|\psi^{grwa}_{-,N}\rangle\langle\psi^{grwa}_{+,N}|$, 
and 
\begin{equation} \label{MqNp}
\ket{M^{(q_N)}_{\pm^\prime}} =  D(q_N/\omega_m) D(\mp g^{(N)}_{eff} /\omega_m) \ket{M}.
\end{equation} 
We added a prime label to emphasize the phononic nature of these
displaced states in comparison with the displaced photonic states $\ket{N_{\pm}}$
(see Eq.(\ref{N+-})). The displacement of $\ket{M}$ by $q_N$ arises from the introduction of the displaced phonon operators $b_N$, while the shift by $\pm g^{(N)}_{eff}/\omega_m$ 
is a consequence of a further GRWA treatment of the Rabi Hamiltonian (\ref{hibi})
(``GRWA-GRWA'' approach). 
The isolated state (\ref{2323232}) contains $N_R^{grwa}=N+1$ atom-cavity polaritons and, unlike those of a RWA-RWA treatment \cite{PhysRevA.95.023832}, is not separable. 
The corresponding energy is
\begin{equation}
E^{(N)}_G= C_N - \frac{({g}_{eff}^{(N)})^2}{\omega_m} - \frac{T_{N}}{2} \braket{ (M=0)^{(q_N)}_{-^\prime}  | (M=0)^{(q_N)}_{+^\prime}} .
\label{grwadim1}
\end{equation}
Note that the last term introduces a dependence on the coupling parameters
$g_{ac}$ and $ g_{om}$, in remarkably contrast to the RWA approach.
Figure \ref{plot_energias}(b) displays the function $E^{(N)}_G(g_{ac})$ for $N=0,1,...,5$ at resonance $\omega_a=\omega_c$. 
After an initial agreement between all the calculations for small 
coupling, only the GRWA curve (\ref{grwadim1}) accurately follows the exact results for
increasing coupling strength. The RWA-RWA presents a linearly decreasing behavior
without any curvature, which it is apparent in the numerical calculation.
It is clear how the GRWA method describes very precisely the energies even in the deep strong coupling. We also calculated some off-resonant $\omega_c > \omega_a$ cases
(not shown) and found that the deviation of the RWA-RWA results start
for even smaller values of the coupling.

Finally, the spectrum of dressed polariton-phonon states, arising from the doublets structure in $H^{(N)}$ (\ref{hibi}), are given by
\begin{equation}
E^{(N)}_{\pm,M} =   C_N +  \omega_m \left(M + \frac{1}{2} \right) -\frac{(g_{eff}^{(N)})^2}{\omega_m} + \frac{1}{4}(\Omega^\prime_{M,M}+\Omega^{\prime N}_{M+1,M+1}) \pm \frac{1}{2}\bigl({\Delta^\prime}_{N,M}^2  +  (\Omega^{\prime N}_{M,M+1})^2 \bigl)^{1/2}.
\label{iniciogrwa}
\end{equation}
Motivated by the GRWA of the QRM, 
we wrote these energies in terms of 
a polariton-phonon Rabi frequency 
and a polariton-phonon detuning, defined respectively as
\begin{eqnarray}
{\Omega^{\prime}}^{N}_{M,M+1} &=& T_{N} \braket{M^{(q_N)}_{-^{\prime}}|M+1^{(q_N)}_{+^{\prime}}} \\
\Delta^{\prime}_{NM} &=& \frac{T_N}{2} \bigl( \braket{M^{(q_N)}_{-^{\prime}}|M^{(q_N)}_{+^{\prime}}} + \braket{M+1^{(q_N)}_{-^{\prime}}|M+1^{(q_N)}_{+^{\prime}}}  \bigl) - \, \omega_m \ .
\end{eqnarray}
\begin{figure}
\centering
\includegraphics[width=8.9cm]{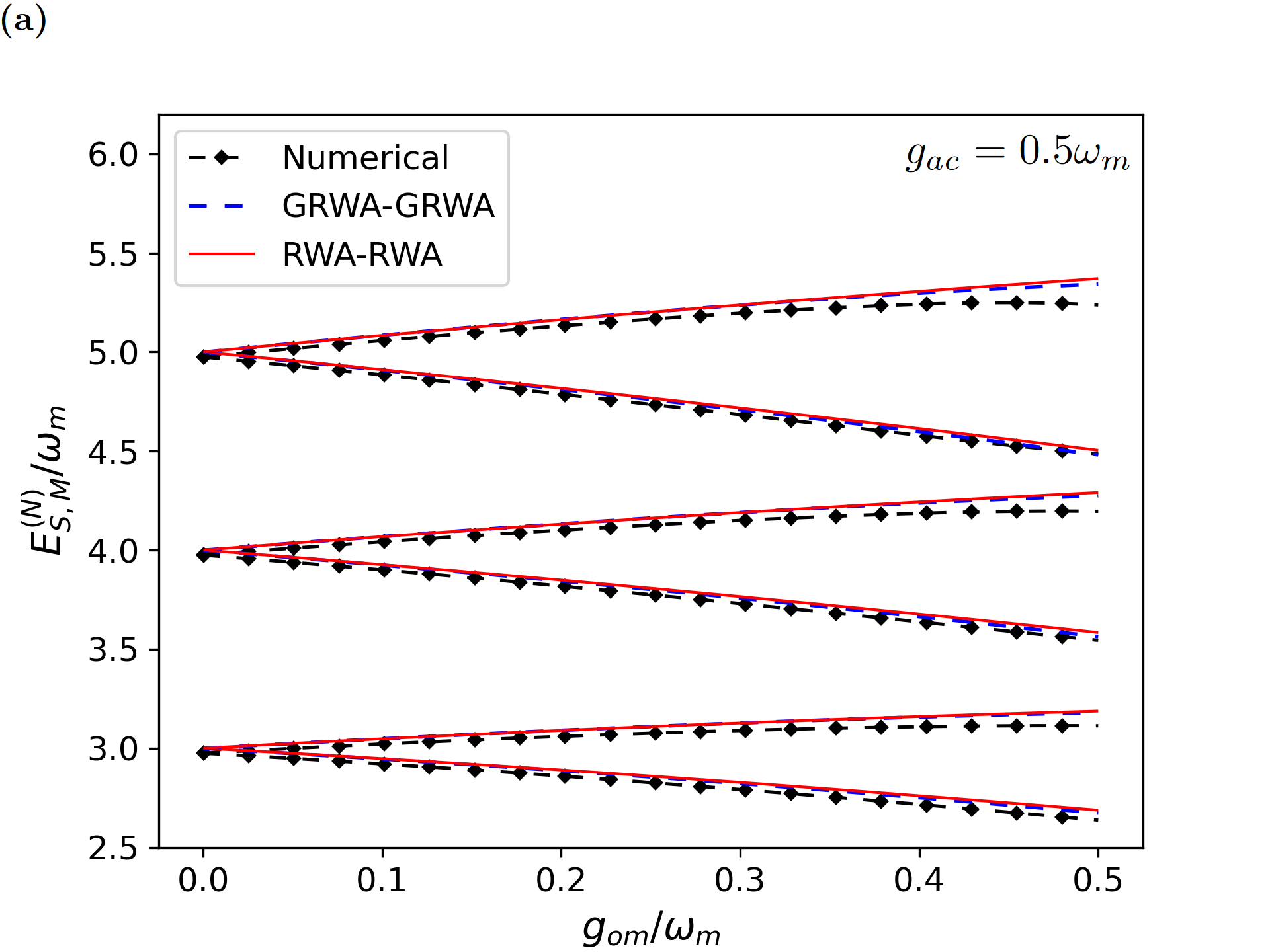}\includegraphics[width=8.9cm]{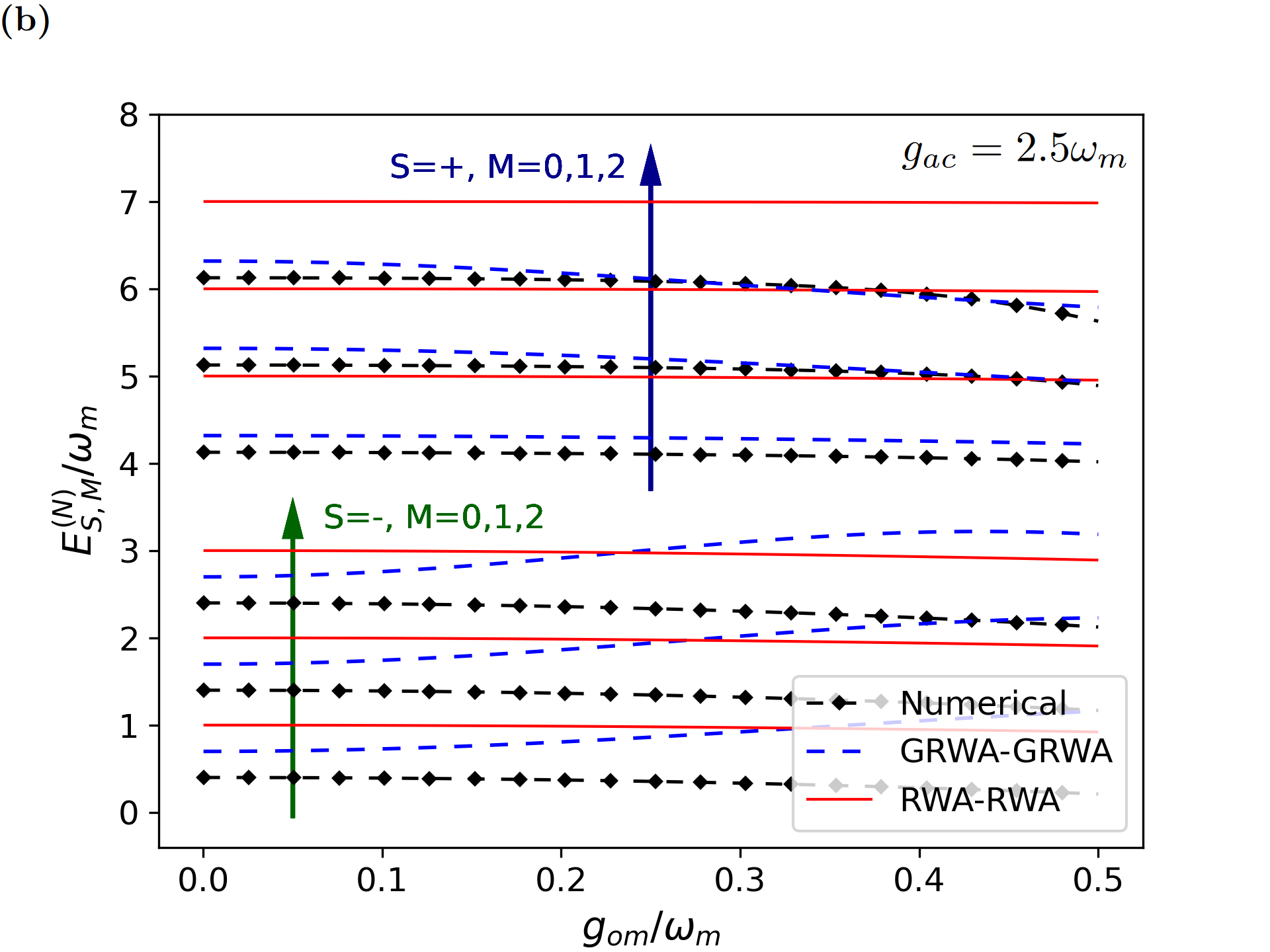}
\justifying
\caption{Low energy levels $E^{(N)}_{\pm,M}$ versus the optomechanical coupling
$g_{om}$, at resonance $\omega=\omega_c=5\,\omega_m$, within RWA (red solid line) and GRWA (blue dashed line) procedures, 
and the numerical exact solution (black dotted-dashed line). (a) $g_{ac}=0.5\,\omega_m$.
(b) $g_{ac}=0.5\,\omega_m$. In (a) the curves with positive (negative) slopes correspond to $S=+\,(S=-)$.}
 \label{g2yg3}
\end{figure}
The eigenstates read as
\begin{eqnarray}
\ket{\Psi^{(N)}_{+,M}} &=& \sin (\phi^{(N)}_{M}/2) \ket{\Psi^{ad}_{+,N,M}} + \cos (\phi^{(N)}_{M}/2) \ket{\Psi^{ad}_{-,N,M+1}}, 
\label{dressed+} \\
\ket{ \Psi^{(N)}_{-,M} } &=& \cos (\phi^{(N)}_{M}/2)  \ket{\Psi^{ad}_{+,N,M}} - \sin (\phi^{(N)}_{M}/2)  \ket{\Psi^{ad}_{-,N,M+1}},
\label{dressed-}
\end{eqnarray}
which are written in terms of the hybrid adiabatic basis
\begin{equation}
\ket{\Psi^{ad}_{\pm,N,M}} = \bigl( \ket{\psi^{grwa (x)}_{+,N}} \ket{M^{(q_N)}_{+^{\prime}} } \pm  \ket{\psi^{grwa (x)}_{-,N}} \ket{M^{(q_N)}_{-^{\prime} } }\bigl)/\sqrt{2}.
\end{equation}
The angle $\phi^{(N)}_{M}$ is defined by
$\tan{\phi^{(N)}_{M}} = - {\Omega^{\prime}}^{N}_{M,M+1}/\Delta^{\prime}_{NM}$.

The energy levels $E^{(N)}_{\pm,M}$, at resonance $\omega_a=\omega_c$ and out
of resonance $\omega_c>\omega_a$, are shown in figures
\ref{plot_energias}(c) and \ref{plot_energias}(d), respectively. 
The GRWA-GRWA energies (\ref{iniciogrwa}) reproduce very well the
exact (numerical) calculation in the whole range of couplings $g_{ac}$.
The discrepancy of the RWA-RWA curves is clearly visible.

Figures \ref{plot_energias}(e) and \ref{plot_energias}(f) show the
overlap $| \braket{\Psi^{(0)}_{\pm,2} |\Psi^{numerical}}|^2$ 
between the GRWA-GRWA eigenstates $\ket{\Psi^{(0)}_{\pm,2}}$ and their numerical counterpart, under resonant and off-resonant conditions respectively, in comparison
with the RWA-RWA result. Within the GRWA, the fidelity lies above 80\%, obtaining the highest value for the state $\ket{\Psi^{(0)}_{-,2}}$, while the RWA fidelity shows
a departure from the optimal coincidence for increasing values of coupling $g_{ac}$.

The behavior of the spectrum $E^{(N)}_{\pm,M}$ as a function of the optomechanical coupling $g_{om}$, at resonance $\omega_c=\omega_a$, is shown in Fig.\,\ref{g2yg3}. 
At small $g_{ac}=0.5 \, \omega_m$ (Fig.\,\ref{g2yg3}(a)), both approximations display
a good agreement with the exact results. At higher values, $g_{ac}=2.5\,\omega_m$,
the RWA-RWA curves clearly fail, while the GRWA-GRWA energies with $S=+1$ follow closely the exact results (Fig.\,\ref{g2yg3}(b)). Note however that the branches with
$S=-1$ show a moderate discrepancy. 
On the other hand, under the off-resonance condition $\omega_c>\omega_a$,
all the spectra (not shown) $E^{(N)}_{\pm,M}$, $E_M$, and $E^{(N)}_G$, approximate  the numerical solution much better.

In Ref.\,\onlinecite{PhysRevLett.99.173601} the limited accuracy of the GRWA of the QRM for large positive
detuning $\omega_a-\omega_c$ ($\sim\omega_c$) was discussed. By analogy,
from (\ref{hibi}) one could expect a limited precision for large $T_N-\omega_m$. 
The increase of the $g_{ac}$ coupling implies the increase of the generalized
GRWA Rabi frequency $T_N$ (see Eq.(\ref{GTN})) in the range considered in Fig.\,\ref{plot_energias}, and the fairly good agreement
observed in Fig.\,\ref{plot_energias}(c) means that the GRWA-GRWA approach works
for a wider range of positive detuning $T_N-\omega_m$.

\section{Bipartite entanglement}

In this section we address the entanglement properties of the GRWA states
(\ref{dressed+}) and (\ref{dressed-}). The reduction of the hybrid Hamiltonian
(\ref{full_hyb}) to an effective spin-boson Hamiltonian (\ref{hibi}), 
which describes a two polariton level system interacting with 
a displaced phononic field, allows to study the entanglement as that of a
bipartite system. Thus, the Schmidt number criterion can be applied to 
reveal the non-separability of the states \cite{Giuliano}. To this end, we calculate the participation ratio $\xi=1/\text{Tr}(\hat{\rho}^2)$ 
in order to quantify the degree of entanglement of the dressed GRWA states, where $\hat{\rho}$ is a reduced density matrix obtained by a partial trace on any of the subsystems. The tripartite entanglement properties of a JCM with optomechanical interaction was recently studied, although in the resonant weak coupling regime only \cite{Liao2018}.

For the sake of completeness, we first consider the QRM states $|\psi^{grwa}_{\pm,N}\rangle$ (\ref{psigrwa+}) and (\ref{psigrwa-}).

\subsection{Entanglement of QRM states}

The density matrices $\hat{\rho}^{grwa}_{\pm,N}=\ket{\psi^{grwa}_{\pm,N}} \bra{\psi^{grwa}_{\pm,N}}$ of the GRWA Rabi Hamiltonian $\hat{H}^{grwa}_R$ 
(\ref{HGRWA}) are given by  
\begin{eqnarray}
\label{density_grwa}
    \hat{\rho}^{grwa}_{\pm,N} &= & f_{\pm}^{2}(\alpha_N/2) \ket{\psi^{ad}_{+,N}}\bra{\psi^{ad}_{+,N}} + f_{\mp}^{2}(\alpha_N/2) \ket{\psi^{ad}_{-,N+1}}\bra{\psi^{ad}_{-,N+1}}  \\&  & \hspace*{5cm} \pm  f_{\pm}(\alpha_N/2) f_{\mp}(\alpha_N/2) \bigl( \ket{\psi^{ad}_{+,N}}\bra{\psi^{ad}_{-,N+1}} +\ket{\psi^{ad}_{-,N+1}}\bra{\psi^{ad}_{+,N}} \bigl). \nonumber
\end{eqnarray}
where $f_{+}(x) = \sin x$ and $f_{-}(x) = \cos x$, such that
$f^2_{\pm}(\alpha_n/2)=(T_N\pm\Delta_N)/2T_N$, and $\ket{\psi^{ad}_{\pm,N}}$
are the adiabatic states (\ref{psiad}).
Taking the trace over the photonic subsystem, we obtain the reduced density matrices
\begin{equation}
    \tilde{\hat{\rho}}^{grwa}_{\pm,N} = \frac{1}{2} \bigl[ \ket{+x}\bra{+x} + \ket{-x}\bra{-x} + \lambda_{\pm,N} \bigl( \ket{+x}\bra{-x} + \ket{-x}\bra{+x} \bigl) \bigl]
    \label{ro_red_grwa}
\end{equation}
where 
\begin{equation}
    \lambda_{\pm,N} = f^2_{\pm}(\alpha_N/2) \braket{N_-|N_+} - f^2_{\mp}(\alpha_N/2) \braket{N+1_-|N+1_+} \pm 2 f_{\pm}(\alpha_N/2)f_{\mp}(\alpha_N/2) \braket{N_-|N+1_+}.
\end{equation}
The corresponding participation ratio is
\begin{equation}
    \xi_{\pm,N}^{grwa} = \frac{2}{1 + \lambda^2_{\pm,N} }.
    \label{grwa_schmidt}
\end{equation}

This expression resembles the participation ratio of the Jaynes-Cummings states 
\begin{equation}
\ket{\psi^{JC}_{\pm,N}}=f_{\pm}(\beta_N/2) \ket{+z,N} \pm f_{\mp}(\beta_N/2) \ket{-z,N+1},
\end{equation}
which reads as
\begin{equation}
\xi^{rwa}_{N} = \frac{2}{1 + \cos^2\beta_N },
\label{jc_schmidt}
\end{equation}
where $\tan\beta_N= -2g_{ac}\sqrt{N+1}/(\omega_a-\omega_c)$.

\begin{figure}[ht]
\centering
\includegraphics[width=8.9cm]{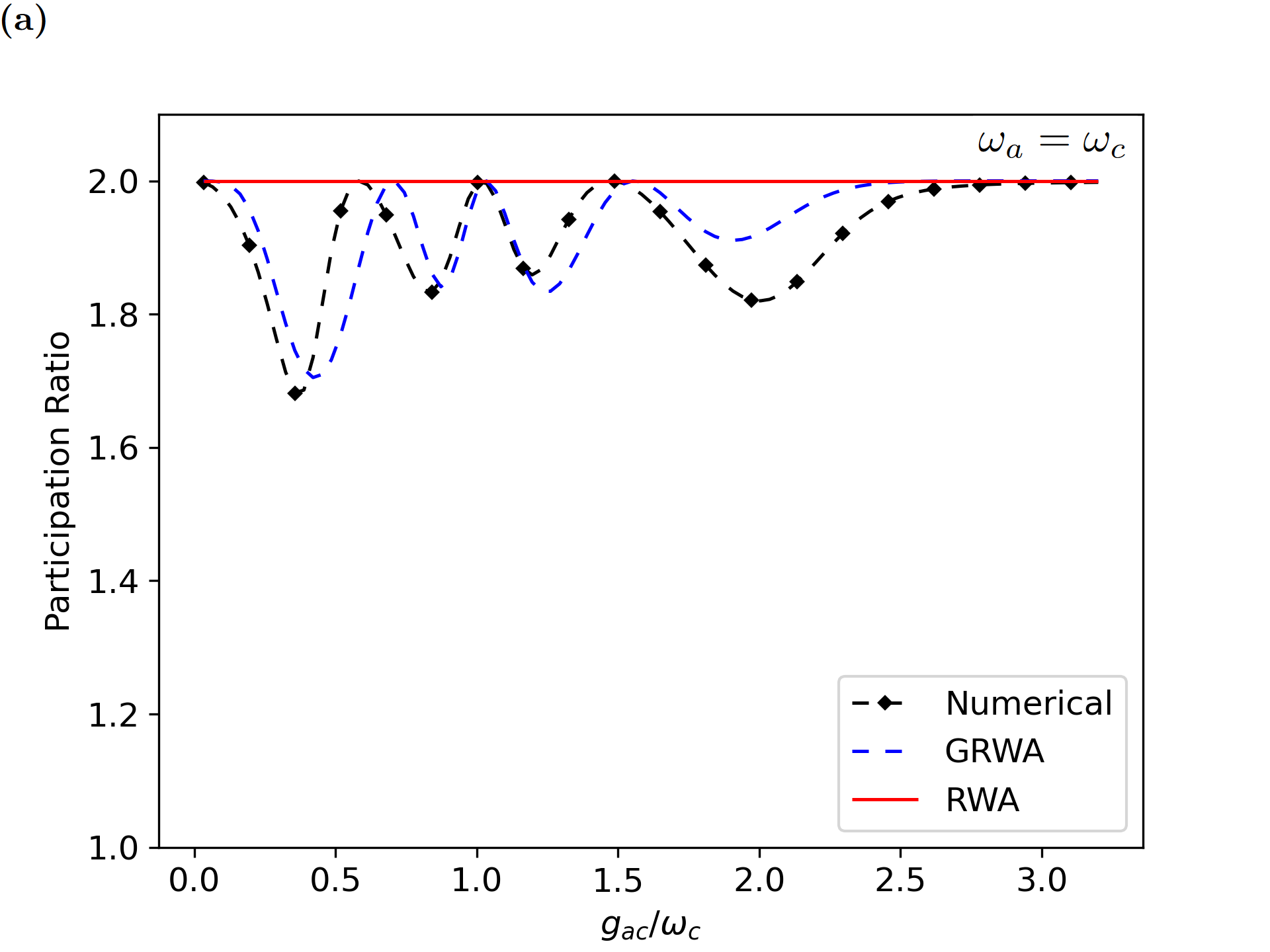}\includegraphics[width=8.9cm]{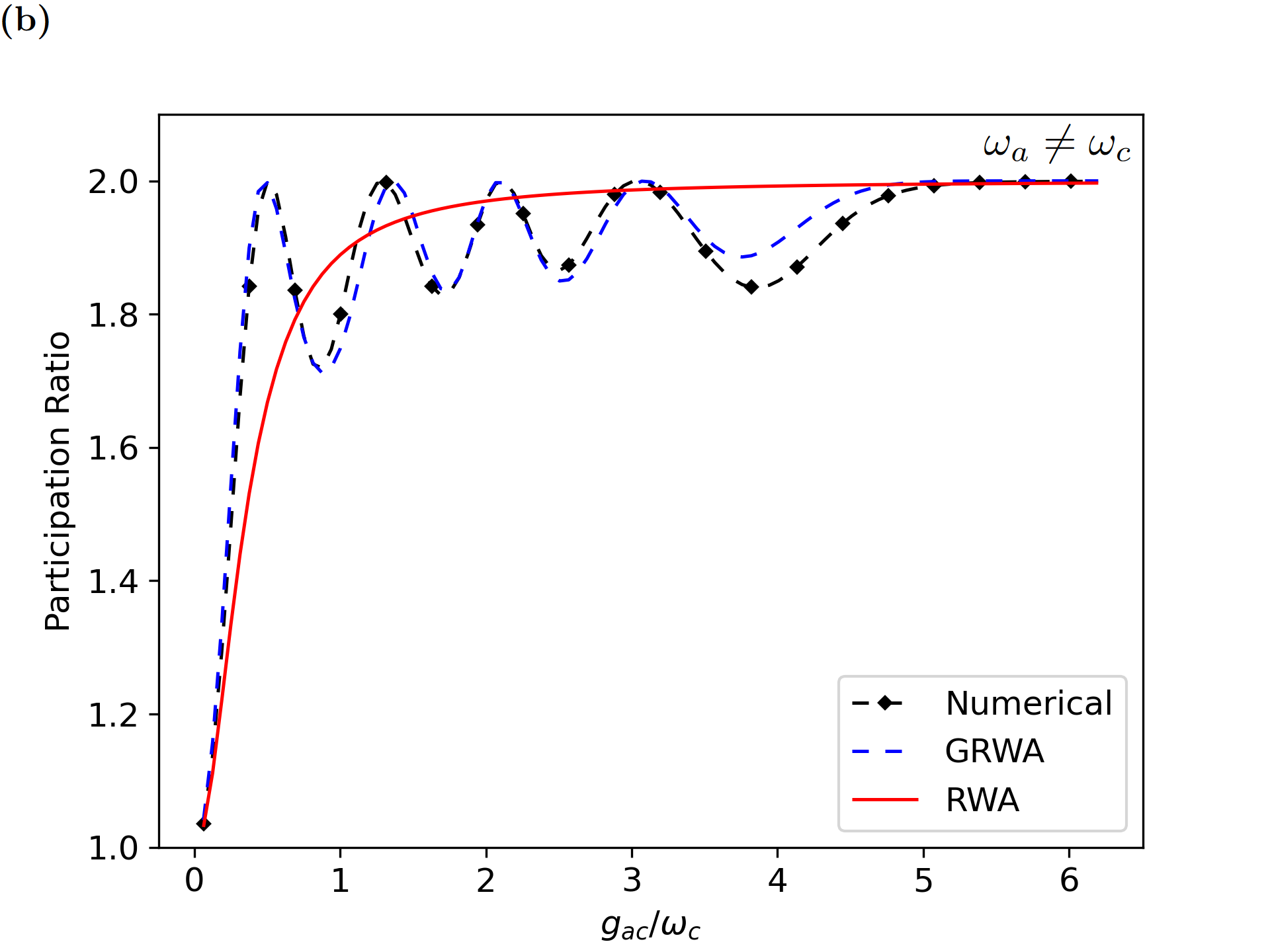}
\justifying
\caption{Participation ratio as a function of the atom-cavity coupling $g_{ac}$ 
for the Jaynes-Cummings state $|\psi^{JC}_{+,3}\rangle$, the GRWA state 
$|\psi^{grwa}_{+,3}\rangle$, and the corresponding exact (numerical) result. 
(a) Resonant case $\omega_a = \omega_c$, (b)  off-resonant case 
$\omega_c=2 \omega_a$.}
\label{sch_rabi}
\end{figure}

Figure \ref{sch_rabi} shows a comparison between the results
(\ref{grwa_schmidt}), (\ref{jc_schmidt}), and the numerical, for a particular state. 
The participation ratio of the JC states predicts maximum entanglement for 
any coupling value, under resonance conditions $\omega_c=\omega_a$
(Fig. \ref{sch_rabi}(a)). In contrast, the GRWA states present an oscillatory behavior following that of the numerical solution. 
For the off-resonant case $\omega_c=2\omega_a$ (Fig.\,\ref{sch_rabi}(b)),
now $\xi^{rwa}_{N}$ increase monotonically, suggesting increase of the entanglement.
However, $\xi_{\pm,N}^{grwa}$ still behaves non-monotonically, fitting the
numerical result closely. The observed oscillations of the GRWA result arise from the 
overlap between displaced photonic states (\ref{overlapMN}). 

\subsection{Entanglement of hybrid states}

We consider now the entanglement between the atom-cavity polaritons and phonons. 
The density matrix for the dressed states (\ref{dressed+}) and (\ref{dressed-}) is
\begin{equation}
\hat{\rho}^{(N)}_{\pm,M} = \ket{\Psi^{(N)}_{\pm,M}}\bra{\Psi^{(N)}_{\pm,M}}. 
\label{rhodressed+}
\end{equation}
Taking the trace over the phononic subsystem, we obtain the reduced density operator
\begin{equation}
\Tilde{\hat{\rho}}^{(N)}_{\pm,M} = \frac{1}{2} \bigl[   \ket{\psi^{grwa (x)}_{+,N}} \bra{\psi^{grwa (x)}_{+,N}} +\ket{\psi^{grwa (x)}_{-,N}} \bra{\psi^{grwa (x)}_{-,N}}   +\Lambda^{(N)}_{\pm,M} \bigl( \ket{\psi^{grwa (x)}_{+,N}} \bra{\psi^{grwa (x)}_{-,N}} + \ket{\psi^{grwa (x)}_{-,N}} \bra{\psi^{grwa (x)}_{+,N}} \bigl) \bigl] 
\label{tdressed+}
\end{equation}
with 
\begin{eqnarray}
\Lambda^{(N)}_{\pm,M} &=&  f^2_{\pm}(\phi^{(N)}_{M}/2) \braket{M^{(q_N)}_{-^{\prime}}|M^{(q_N)}_{+^{\prime}} } -  f^2_{\mp}(\phi^{(N)}_{M}/2) \braket{M+1^{(q_N)}_{-^{\prime}}|M+1^{(q_N)}_{+^{\prime}}} \\
&& \hspace*{6cm} \pm 2 f_{\pm}(\phi^{(N)}_{M}/2) f_{\mp}(\phi^{(N)}_{M}/2) \braket{M^{(q_N)}_{-^{\prime}} |M+1^{(q_N)}_{+^{\prime}}}, \nonumber
\label{Lambda_nm}
\end{eqnarray}
where the angle $\phi^{(N)}_{M}$ is defined below the states (\ref{dressed+}) and 
(\ref{dressed-}), and the states $\ket{M^{(q_N)}_{\pm^\prime}}$ in (\ref{MqNp}).  

This leads to the following participation ratio,
\begin{equation}
\xi^{(N)}_{\pm,M} = \frac{2}{1 + {[\Lambda^{(N)}_{\pm,M}}]^2}\ .
\label{sch_exp1}
\end{equation}
\begin{figure}[ht]
\centering
\includegraphics[width=8.9cm]{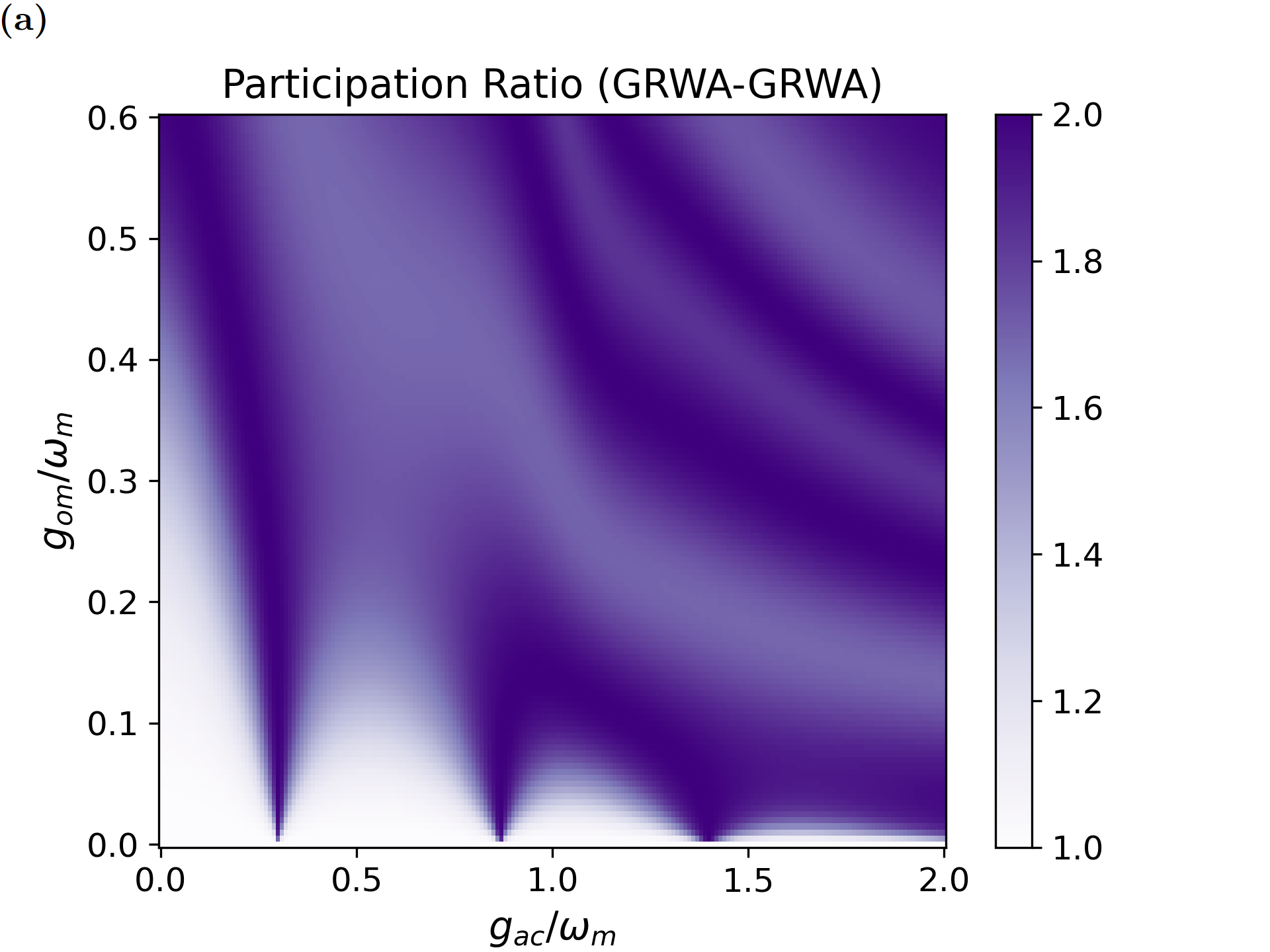} \includegraphics[width=8.9cm]{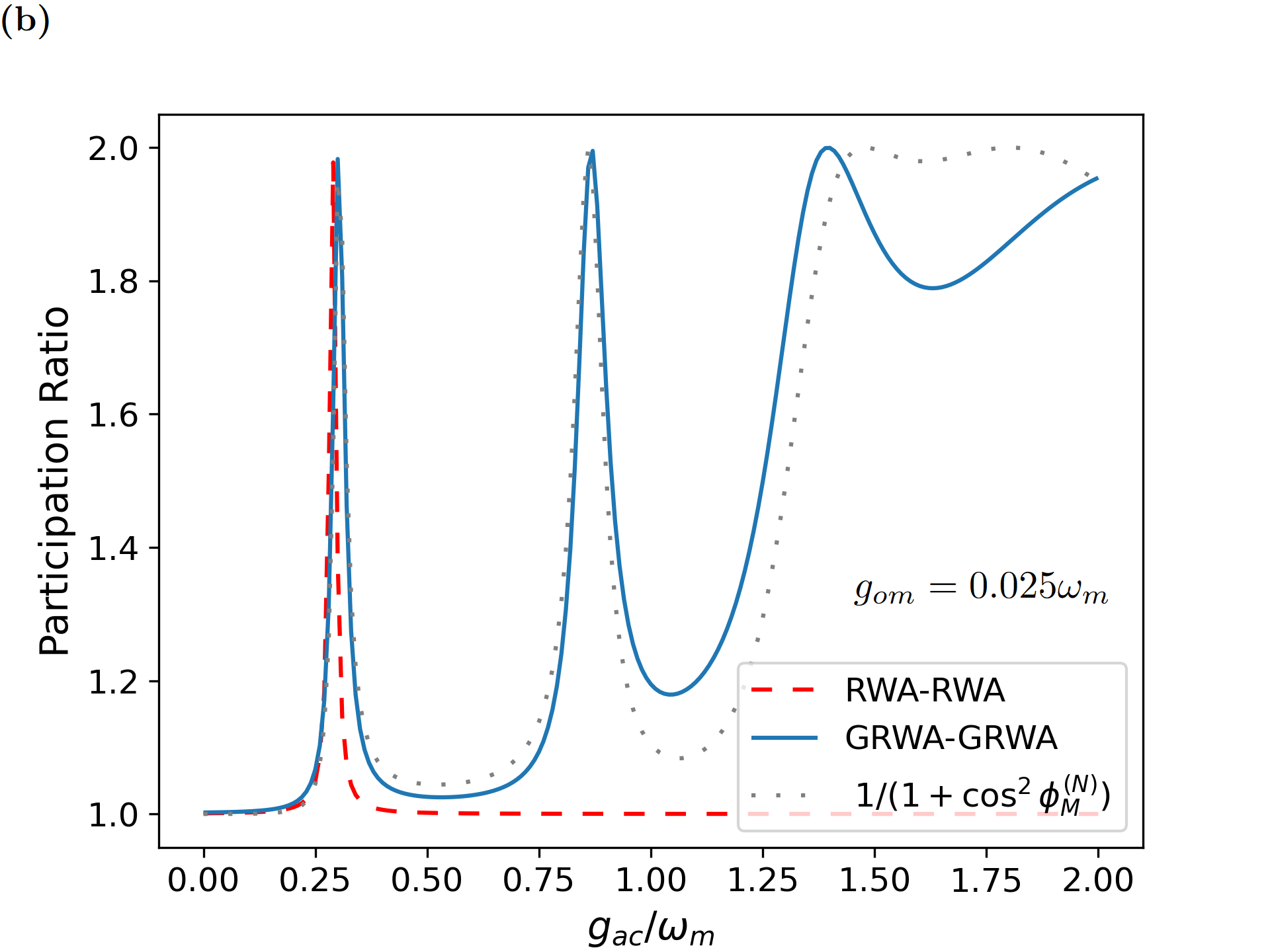}
\includegraphics[width=8.9cm]{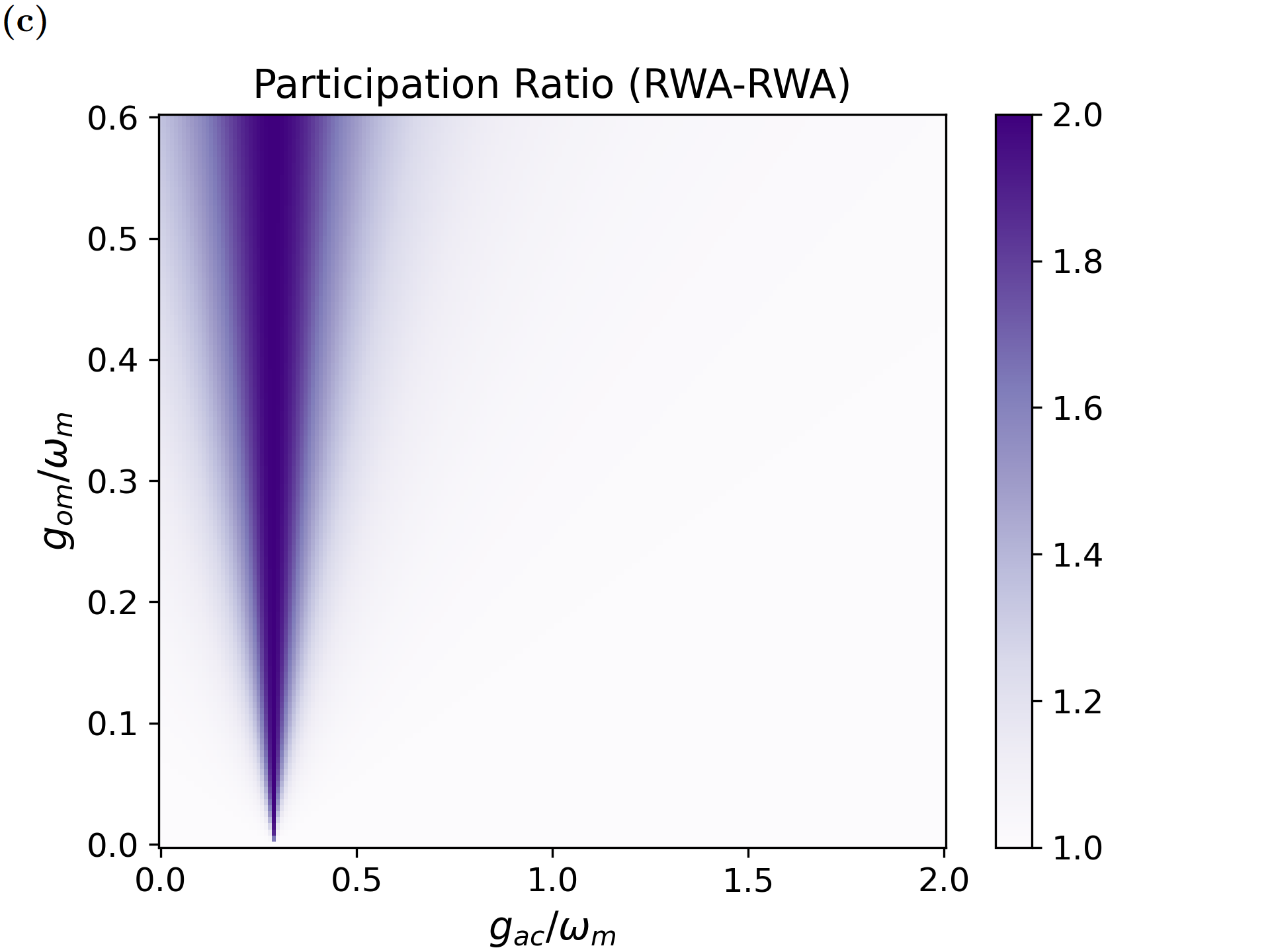}\includegraphics[width=8.9cm]{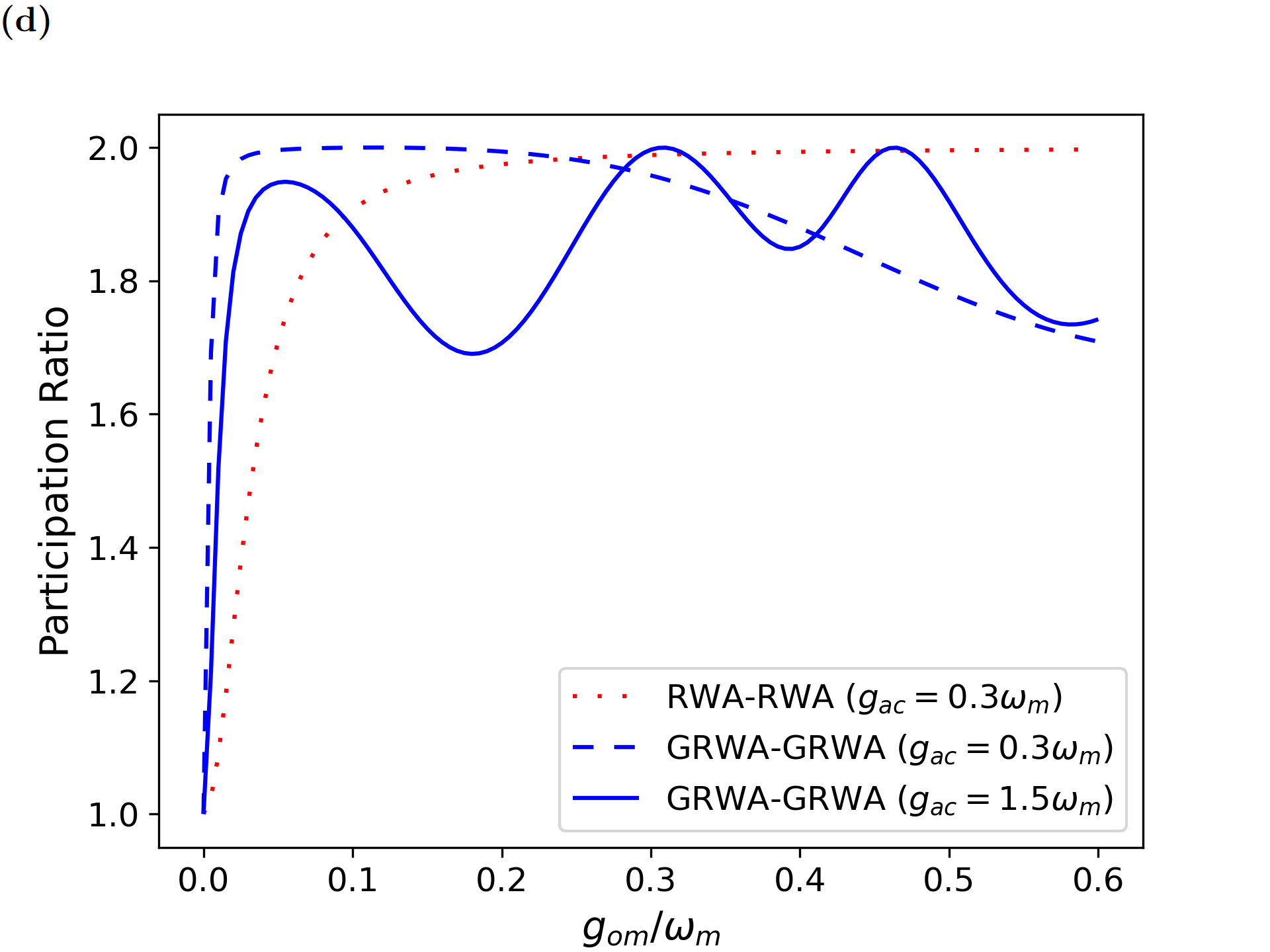}
\justifying
\caption{(a) Participation ratio $\xi^{(2)}_{+,2}$ of the GRWA state $\ket{\Psi^{(2)}_{+,2}}$ as a function of the optomechanical and atom-cavity 
couplings of the hybrid system. (b) The numbers $\xi^{(2)}_{+,2}$ and 
$\xi^{rwa}_{2,2}$ as a function of $g_{ac}$ for a small value of $g_{om}$.
(c) Participation ratio $\xi^{rwa}_{2,2}(g_{ac},g_{om})$ of the corresponding hybrid 
RWA state. (d) The function $\xi^{(2)}_{+,2}$ versus optomechanical coupling, for
two values of the coupling strength $g_{ac}$.}
\label{plot_sch}
\end{figure}

The same procedure for the hybrid eigenstates within the RWA-RWA approach gives
\begin{equation}
\xi^{rwa}_{N,M} = \frac{2}{1 + \cos^2\theta^{(N)}_M  }, 
\label{sch_exp2}
\end{equation}
with the angle $\theta^{(N)}_M$ defined by
$\tan \theta^{(N)}_M = - g_{om}\sqrt{M+1}/(R_N - \omega_m)$,
where $R_N = \sqrt{(\omega_a-\omega_c)^2+\Omega_N^2} $ is the well known generalized Rabi frequency of the Jaynes-Cummings model.

Figure \ref{plot_sch}(a) shows a color map of $\xi^{(N)}_{\pm,M}(g_{ac},g_{om})$ (\ref{sch_exp1}) for the eigenstate $\ket{\Psi^{(2)}_{+,2}}$ at resonance $\omega_a =\omega_c = \omega_m$. 
For small values of $g_{om}$, the polariton-phonon entanglement occurs maximally
for a set of narrow ranges of $g_{ac}$. For increasing magnitude of the optomechanical coupling the structure widens, displaying a larger region of values $(g_{ac},g_{om})$ with maximal entanglement $\xi^{(2)}_{+,2}\approx 2$. 
In order to explain this structure, we show in Fig.\,\ref{plot_sch}(b) a cut
of the map with the line $g_{om}=0.05\omega_m$. The participation ratio
$\xi^{(2)}_{\pm,2}$ presents three peaks in the whole range of $g_{ac}$, from
the weak coupling to the deep strong coupling regime. This behavior can be
understood by noting that when $g^{(N)}_{eff}/\omega_m\ll 1$, then
$\braket{M^{(q_N)}_{-^{\prime}} |M+1^{(q_N)}_{+^{\prime}}}\approx 0$, 
$\braket{M^{(q_N)}_{-^{\prime}}|M^{(q_N)}_{+^{\prime}} }\approx 1$, $\braket{M+1^{(q_N)}_{-^{\prime}}|M+1^{(q_N)}_{+^{\prime}}}\approx 1$ (see Eq.(\ref{MqNp})),
and therefore $\Lambda^{(N)}_{\pm,M}\approx \pm \cos \phi^{(N)}_M$,
$\xi^{(N)}_{\pm,M}\approx 2/[1+\cos^2\phi^{(N)}_M]$ 
(dashed line in Fig.\,\ref{plot_sch}(b)). It can be verified that
the term $\cos\phi^{(2)}_2$ develops a structure with three minima, arising from a
combination of the
Laguerre polynomials $L_2^0(x), L_3^0(x), L_2^1(x)$ with 
$x=2g^{(2)}_{eff}/\omega_m$, where $g^{(2)}_{eff}$ involves in turn
the coupling $g_{ac}$ and the photonic 
overlaps $\langle 2_-|2_+\rangle, \langle 3_-|3_+\rangle$, and $\langle 2_-|3_+\rangle$
(see (\ref{overlapMN})), through the frequency $\Omega_N$ and the
angle $\alpha_2$ given by
$\tan\alpha_2=-\Omega_{2,3}/\Delta_2$. For increasing $g^{(2)}_{eff}$
(or $g_{om}$), the phononic
overlaps in $\Lambda^{(2)}_{+,2}$ (\ref{Lambda_nm}) start to deviate from trivial values
(0 or 1), leading to the more structured map observed in 
Fig.\,\ref{plot_sch}(a), where $\xi^{(2)}_{+,2}(g_{ac},g_{om})\approx 2$.
We can explain the behavior of the map shown in Fig.\,\ref{plot_sch}(a) as
arising from the photonic overlaps and phononic overlaps.

In contrast, the corresponding RWA-RWA participation ratio
$\xi^{rwa}_{N,M}(g_{ac},g_{om})$ (\ref{sch_exp2}) develops only one 
triangular structure with maximal entanglement, as is illustrated
in Fig.\,\ref{plot_sch}(c). For a fixed value of $g_{om}$, only one
hump appears as a function of $g_{ac}$. This is illustrated in 
Fig.\,\ref{plot_sch}(b) for small optomechanical coupling.
In this case, the peak arises from the single zero of $\cos^2\theta^{(2)}_2$ 
(see (\ref{sch_exp2})) at $\Omega_2/\omega_m=1$ (i.e. $g_{ac}/\omega_m=\sqrt{3}/6$ in
the figure).
The results $\xi^{(N)}_{\pm,M}$ and $\xi^{rwa}_{N,M}$
match for weak coupling, as expected,  but for strong coupling strengths only the 
GRWA-GRWA approach predicts maximum entanglement.
 The slight discrepancy between the first peaks predicted by GRWA and RWA calculations becomes more apparent when the frequency $T_N$ differs appreciably from $R_N$.

In Figure \ref{plot_sch}(d) a cut of the map with the lines $g_{ac}=0.3\omega_m$
and $g_{ac}=1.5\omega_m$ are shown. For small $g_{ac}$ the RWA result grows
monotonically and saturates close to $\xi^{rwa}_{2,2}\approx 2$, while 
$\xi^{(2)}_{+,2}$ starts to develop a hump. For the larger value of $g_{ac}$,
within the deep-strong-coupling regime, now $\xi^{(2)}_{+,2}$ displays 
some oscillations which start for lower values of the optomechanical coupling.

A remaining part of the spectrum is the isolated state $\ket{\Psi_G^{(N)}}$ (\ref{2323232}), for which 
\begin{equation}
    \xi^{(N)}_{G} = \frac{2}{1 + [\braket{ (M=0)^{(q_N)}_{-^{\prime}} | (M=0)^{(q_N)}_{+^{\prime}} }]^2 }.
    \label{P_Ratio_G}
\end{equation}
For small (large) effective coupling $g^{(N)}_{eff}$,
the overlap between oppositely displaced phonon states is close to one (zero) and thus $\xi^{(N)}_G\approx 1\,(0)$, that is, the state $\ket{\Psi_G^{(N)}}$
begins to be non-separable (separable). The corresponding RWA-RWA state exhibits
an absence of entanglement for any value of $g_{ac}$ and $g_{om}$.

It is clear that the zero atom-cavity polaritons states $\ket{\Psi_M}$ (\ref{finalgrwa}) are separable and thus there will be no polariton-phonon 
entanglement in both approaches, $\xi_{M}=1$. 

\section{Summary}

Following the strategy of an approximation developed to be valid  
in the strong coupling regime of the quantum Rabi model, we apply
a generalized rotating wave approximation (GRWA) to an  
atom-photon-oscillator system. This hybrid model was introduced  
by Restrepo et al. \cite{PhysRevLett.112.013601}, with the atom-photon interaction
given by the Jaynes-Cummings model, which assumes weak 
coupling and quasi-resonance (RWA).

We found that GRWA approach allows to break the full hybrid Hamiltonian 
into an infinite set of $2\times 2$ blocks, each one described by a 
spin-boson model and characterized by a composed conserved quantity. 
Now the role of the two-level atom is played by the dressesed atom-photon
states (polaritons) and the bosonic part by the quantized mechanical modes.
This effective JCM is treated in turn by another GRWA, in order to go beyond the
RWA conditions. Through this double approximation, we obtained analytical expressions for the energies which show
a very good agreement with the numerical exact solution in a wide range of
atom-photon couplings, from the weak to the deep-strong-coupling regime, and for
large detunings. Importantly, the use of the adiabatic basis states 
involve the interference between oppositely displaced photonic states and 
between displaced phononic states. As a consequence,
the effective couplings, the generalized Rabi frequencies and detunings introduce
oscillations in the energies, which follow closely those of the numerical solution. These oscillations are absent in the RWA-RWA treatment 
of the hybrid model\cite{PhysRevLett.112.013601}.
On the other hand, as a function of the photon-phonon coupling, the agreement
is moderate for large values, but still much better than the RWA-RWA result.
Similarly, the calculated eigenstates also demonstrate the good quality of the GRWA-GRWA approach, showing good fielity when overlaped with the numerical states. 
Because of the analytical character of the approach,
we were able to obtain also a closed form of the participation ratio of the 
GRWA-GRWA states of the bipartite Hamiltonian. This quantity displays a 
 structure with maxima and minima, in strong contrast to the monotonic
behavior of the RWA-RWA results. The participation ratio predicts polariton-phonon entanglement in coupling regions where the solution based on the RWA does not.

A number of experiments on a variety of settings have demonstrated the
emergence of novel optical phenomena in the ultrastrong and deep-strong 
coupling regime of light-matter interaction, as described by the quantum Rabi model \cite{RevModPhys.91.025005}. Thus, in a quantum 
platform combining cavity QED and cavity optomechanics like that studied here, new effects can be expected in the large coupling regime of the atom-photon-phonon
interaction. Our work might be useful about this.  Geometric phases, quantum correlations, quasiparticles statistics, polariton-assisted cooling of the mechanical motion, among other properties of the quantum dynamic of the strongly coupled hybrid system, are some of the phenomena that could be investigated, taking our work as a basis. We hope it encourages further studies.

\section*{Acknowledgments}

W.H.M. acknowledges support from CONACyT (M\'exico). The authors acknowledge 
support from Proyecto DGAPA-PAPIIT IN111122, and thank Fernando Rojas and 
D. Morachis-Galindo for fruitful discussions.

\appendix

 \section{Generalized rotating wave approximation for the Quantum Rabi Model}

The application of the generalized rotating-wave-approximation (GRWA) to the
hybrid Hamiltonian $\hat{H}_{hyb}$ (\ref{full_hyb}) is based on the same strategy used to develop an approximation for the quantum Rabi model in the range of large atom-photon coupling
\cite{PhysRevLett.99.173601}. The sequence of steps in section III employ several basis of states to 
derive analytical expressions of the eigenstates and energy spectrum of the full Hamiltonian (\ref{full_hyb}) within a GRWA approach. In this appendix we review
these basis. The aim is to obtain the GRWA basis $\{|\psi_G^{grwa}\rangle, \ket{\psi^{grwa}_{\pm,N}}, N=0,1,\ldots\}$ used in Eq. (\ref{HGRWA}). To this end, we first need the adiabatic basis \cite{PhysRevB.72.195410} $\{\ket{\psi^{ad}_{\pm,N}}, N=0,1,\ldots\}$, used to rewrite the Rabi Hamiltonian part $\hat{H}_R$, which in turn needs the
displaced oscillator basis $\{|N_{\pm}\rangle, \,N=0,1,\ldots\}$.
\begin{enumerate}
\item {\it Displaced oscillator basis $|N_{\pm}\rangle$.}

This basis is obtained by setting $\omega_a=0$ in the QRM, 
and considering eigenstates of the form $|\pm x\rangle\otimes |N_{\pm}\rangle$
of the Hamiltonian $\hat{H}^{\omega_a=0}_R = \omega_c \hat{a}^\dagger \hat{a} + 
g_{ac} \hat{\sigma}_x (\hat{a}^\dagger + \hat{a})$, 
where $|\pm x\rangle$ are the eigenstates of $\hat{\sigma}_x$. 
Thus, the equation $\hat{H}^{\omega_a=0}_R|\pm x,N_{\pm}\rangle=
E|\pm x,N_{\pm}\rangle$ reduces to
$[\omega_c \hat{a}^\dagger \hat{a}\pm g_{ac}(\hat{a}^\dagger + \hat{a})]
|N_{\pm}\rangle=E|N_{\pm}\rangle$, which corresponds to displaced harmonic
oscillators to the equilibrium positions $\pm 2(g_{ac}/\omega_c)x_0$, where $x_0=\sqrt{\hbar/2m\omega_c}$. The energy spectrum is then given by
\begin{equation}
E_N = \omega_c \bigl(N - g_{ac}^2/\omega_c^2 \bigl), \ \ N=0,1,2,\ldots
\end{equation}
and eigenstates $\ket{\pm x, N_{\pm}}$, with
\begin{equation} \label{N+-}
\ket{N_{\pm}} =  e^{\mp (g_{ac}/\omega_c)(\hat{a}^\dagger- \hat{a}) } \ket{N},
\end{equation}
where $|N\rangle$ are the Fock states of a quantum harmonic oscillator \cite{haroche_book}. Note that the states $|+x,N_+\rangle$ and $|-x,N_-\rangle$ are degenerate
in energy and not mutually orthogonal.

\item {\it Adiabatic Approximation basis $\ket{\psi^{ad}_{\pm,N}}$}. 

The adiabatic approximation for the QRM assumes that the photonic frequency is much larger than the atomic frequency, $\omega_a \ll \omega_c$. In the displaced basis $\ket{\pm x,N_{\pm}}\equiv |s,N_s\rangle$ ($s=\pm$), the restoring of the energy separation $\omega_a\hat{\sigma}_z/2$ in $\hat{H}^{\omega_a=0}_R$ lifts its degeneracy
and introduces non-diagonal terms,
\begin{displaymath}
\langle s',M_{s'}|\hat{H}_R|s,N_s\rangle=
E_N\delta_{s's}\delta_{MN}+(1-\delta_{s's})\langle M_{s'}|N_s\rangle\omega_a/2.
\end{displaymath}
The mixing occurs only between states displaced
oppositely ($s'\neq s$), that is involves only the overlaps $\langle M_{\pm}|N_{\mp}\rangle$.

The adiabatic approximation consists of truncating the matrix to the block diagonal 
form, each block involving levels in opposite wells with the same energy, 
$\langle N_{\pm}|N_{\mp}\rangle$. The assumption $\omega_a \ll \omega_c$
means that a transition in the two-level system can never excite the photonic
field, and in turn means that the mixing occurring between displaced levels with different photon numbers ($M=N$) can be ignored \cite{PhysRevB.72.195410}.
The overlap of two opposite-displaced Fock states is given by
\begin{equation} \label{overlapMN}
    \braket{M_-|N_+} = e^{-2 g_{ac}^2/ \omega^2}(2g_{ac}/\omega)^{N-M} \sqrt{M!/N!}\,
    L^{N-M}_M (4g_{ac}^2/\omega^2), \ \ M \leq N ,
\end{equation}
and $\braket{M_- |N_+  } = (-1)^{M-N} \braket{N_- |M_+  }$,
where $L^{N}_{M}(x)$ is an associated Laguerre polynomial.
The $2\times 2$ block for a given $N$ is given by 
$\hat{H}^{ad}_N=E_N\mathbb{I}+(\omega_a/2)\braket{N_-|N_+}\hat{\sigma}_x$, which
have the spectrum
\begin{eqnarray}
\ket{\psi^{ad}_{\pm,N}} &=& \frac{1}{\sqrt{2}} (\ket{+x,N_+} \pm \ket{-x,N_-}), 
\label{psiad} \\
E^{ad}_{\pm,N} &=& \omega_c \left(N - \frac{g_{ac}^2} {\omega^2_c} \right) \pm  \frac{\omega_a}{2} \braket{N_-|N_+}\,.
\end{eqnarray}
The Laguerre polynomials introduces a rippled structure on the smooth
variation of the energy as a function of $g_{ac}/\omega_c$. It is verified that
$\braket{N_-|N_+}\to 0$ when $g_{ac}/\omega_c\to\infty$, which corresponds to
two uncoupled identical harmonic oscillators and pairwise degenerate
energy levels. The states (\ref{psiad}) appear in the GRWA conserved number $\hat{N}^{grwa}_R$ (\ref{Nnew}). 

\item {\it Generalized rotating-wave approximation basis $|\psi_G^{grwa}\rangle, \ket{\psi^{grwa}_{\pm,N}}$}.

The first step in the derivation of the GRWA is to write the Rabi Hamiltonian
$\hat{H}_R$ in the adiabatic basis (\ref{psiad}), instead of the eigenbasis
$\ket{\pm z, N}$ of the non-interacting Hamiltonian \cite{PhysRevLett.99.173601}. In such a representation,
$\hat{H}_R$ becomes, in matrix form
\begin{equation}
\hat{H}_{R} = \left( \begin{array}{cccccc}

E^{ad}_{-,0} & 0 & 0 & -\frac{1}{2} \Omega_{0,1} & -\frac{1}{2} \Omega_{0,2}  & \cdots \\ [5pt]

0 & E^{ad}_{+,0} & \frac{1}{2} \Omega_{0,1} & 0 & 0 & \cdots \\ [5pt]

0 & \frac{1}{2} \Omega_{0,1} & E^{ad}_{-,1} & 0 & 0 & \cdots \\ [5pt]

-\frac{1}{2} \Omega_{0,1} & 0 & 0 & E^{ad}_{+,1} & \frac{1}{2} \Omega_{1,2} & \cdots \\ [5pt]

-\frac{1}{2} \Omega_{0,2}& 0 & 0 & \frac{1}{2} \Omega_{1,2} & E^{ad}_{-,2} & \cdots \\ 

\vdots & \vdots & \vdots & \vdots & \vdots & \ddots
\end{array} \right).
\label{55.5}
\end{equation}
where $\Omega_{N,N^\prime} = \omega_a \braket{N_- | N^\prime_+}$ is
the atomic frequency renormalized by the overlaps (\ref{overlapMN}). 
The order of columns and rows is $\ket{\psi^{ad}_{-,0}}$, $\ket{\psi^{ad}_{+,0}}$, $\ket{\psi^{ad}_{-,1}}$, $\ket{\psi^{ad}_{+,1}}$, $\ldots$.

The following step is to proceed as the RWA approach for the QRM. This means the 
neglecting of the remote matrix elements (the ``non-conserving energy'', ``counter-rotating'', or ``anti-resonant'' terms), 
which involve two or more excitations, for example 
$\langle\psi_{-,0}^{ad}|\hat{H}_R|\psi^{ad}_{+,1}\rangle=-\Omega_{0,1}/2$.
In terms of the interaction picture, with respect to 
the non-interacting Hamiltonian, the discarded terms are those involving
two or more net excitations, which oscillate very fast.
This reduces the matrix to a 2 × 2 block-diagonal form
\begin{equation}
    H^{grwa}_N = \left(\begin{array}{cccc}
    E^{ad}_{+,N} & \frac{1}{2} \Omega_{N,N+1} \\ [5pt]
    \frac{1}{2} \Omega_{N,N+1} & E^{ad}_{-,N+1} \\ [5pt]
    \end{array}\right).
\end{equation}

Each block constitutes an invariant subspace with a Hamiltonian that can be solved independently, characterized by the conserved quantity (\ref{Nnew}).
 The GRWA energies are
\begin{eqnarray}
E^{grwa}_G &=& E^{ad}_{-,0} = - \frac{g^2_{ac}}{\omega_c} - 
\frac{ \Omega_{0,0} }{2}, \label{g_egrwa} \\
E^{grwa}_{\pm,N} &=&  \omega_c (N + 1/2) - \frac{g_{ac}^2}{\omega_c} \pm \frac{T_N}{2},
    \label{e_egrwa}
\end{eqnarray}
where 
\begin{equation} \label{GTN}
T_N=\sqrt{\Omega^2_{N,N+1} + \Delta_N^2},
\end{equation}
is the generalized GRWA Rabi frequency (see Eq.(\ref{hibi})), 
with  $\Delta_N =  \frac{\omega_a}{2}\left[\braket{N_-|N_+}+\braket{(N+1)_-|(N+1)_+} \right] - \, \omega_c $ being the corresponding generalized detuning.
The corresponding eigenstates are
\begin{eqnarray}
\ket{\psi^{grwa}_G} &=& \ket{\psi^{ad}_{-,0}}, \label{g_sgrwa}\\
\ket{\psi_{+,N}^{grwa}} &=&  \sin(\alpha_N/2) \ket{\psi^{ad}_{+,N}} + \cos(\alpha_N/2) \ket{\psi^{ad}_{-,N+1}} \label{psigrwa+}\\
\ket{\psi_{-,N}^{grwa}} &=&  \cos(\alpha_N/2) \ket{\psi^{ad}_{+,N}} - \sin(\alpha_N/2) \ket{\psi^{ad}_{-,N+1}}\,, \label{psigrwa-}
\end{eqnarray}
where $\tan\alpha_N=-\Omega_{N,N+1}/\Delta_N$. 
\end{enumerate}

The GRWA preserves the accuracy of the RWA for quasi-resonance $\omega_c\approx\omega_a$
and weak coupling conditions $g_{ac}/\omega_c\ll 1$, and that of the adiabatic approximation at large coupling values when $\omega_a\approx 0$. As is explained in Reference \onlinecite{PhysRevLett.99.173601},
the change of basis, from RWA to the adiabatic, retains an important similarity
between the two approximations. In both cases, the calculations start from
a degenerate basis states, the doublet $\{|+z,N\rangle,|-z,N+1\rangle\}$ at resonance
$\omega_c=\omega_a$ with $g_{ac}=0$ in the case of RWA, and 
$\{|+x,N_+\rangle,|-x,N_-\rangle\}$ for $\omega_a=0$, regardless of the value of $g_{ac}$, for the adiabatic. The success of GRWA is due to the fact that takes into account both situations, it works well for all values of the coupling and large negative detuning $\omega_a-\omega_c$, and even for moderate positive detuning $\sim\omega_c$.

The expressions (\ref{g_egrwa})-(\ref{psigrwa-}) define the basis used in the 
spectral decomposition (\ref{HGRWA}), which constitutes the starting point of the GRWA for the full hybrid Hamiltonian $\hat{H}_{hyb}$ (\ref{full_hyb}).

\bibliography{Bibliography}

\begin{thebibliography}{45}%
\makeatletter
\providecommand \@ifxundefined [1]{%
 \@ifx{#1\undefined}
}%
\providecommand \@ifnum [1]{%
 \ifnum #1\expandafter \@firstoftwo
 \else \expandafter \@secondoftwo
 \fi
}%
\providecommand \@ifx [1]{%
 \ifx #1\expandafter \@firstoftwo
 \else \expandafter \@secondoftwo
 \fi
}%
\providecommand \natexlab [1]{#1}%
\providecommand \enquote  [1]{``#1''}%
\providecommand \bibnamefont  [1]{#1}%
\providecommand \bibfnamefont [1]{#1}%
\providecommand \citenamefont [1]{#1}%
\providecommand \href@noop [0]{\@secondoftwo}%
\providecommand \href [0]{\begingroup \@sanitize@url \@href}%
\providecommand \@href[1]{\@@startlink{#1}\@@href}%
\providecommand \@@href[1]{\endgroup#1\@@endlink}%
\providecommand \@sanitize@url [0]{\catcode `\\12\catcode `\$12\catcode
  `\&12\catcode `\#12\catcode `\^12\catcode `\_12\catcode `\%12\relax}%
\providecommand \@@startlink[1]{}%
\providecommand \@@endlink[0]{}%
\providecommand \url  [0]{\begingroup\@sanitize@url \@url }%
\providecommand \@url [1]{\endgroup\@href {#1}{\urlprefix }}%
\providecommand \urlprefix  [0]{URL }%
\providecommand \Eprint [0]{\href }%
\providecommand \doibase [0]{https://doi.org/}%
\providecommand \selectlanguage [0]{\@gobble}%
\providecommand \bibinfo  [0]{\@secondoftwo}%
\providecommand \bibfield  [0]{\@secondoftwo}%
\providecommand \translation [1]{[#1]}%
\providecommand \BibitemOpen [0]{}%
\providecommand \bibitemStop [0]{}%
\providecommand \bibitemNoStop [0]{.\EOS\space}%
\providecommand \EOS [0]{\spacefactor3000\relax}%
\providecommand \BibitemShut  [1]{\csname bibitem#1\endcsname}%
\let\auto@bib@innerbib\@empty
\bibitem [{\citenamefont {Rabi}(1936)}]{PhysRev.49.324}%
  \BibitemOpen
  \bibfield  {author} {\bibinfo {author} {\bibfnamefont {I.~I.}\ \bibnamefont
  {Rabi}},\ }\href {https://doi.org/10.1103/PhysRev.49.324} {\bibfield
  {journal} {\bibinfo  {journal} {Phys. Rev.}\ }\textbf {\bibinfo {volume}
  {49}},\ \bibinfo {pages} {324} (\bibinfo {year} {1936})}\BibitemShut
  {NoStop}%
\bibitem [{\citenamefont {Rabi}(1937)}]{PhysRev.51.652}%
  \BibitemOpen
  \bibfield  {author} {\bibinfo {author} {\bibfnamefont {I.~I.}\ \bibnamefont
  {Rabi}},\ }\href {https://doi.org/10.1103/PhysRev.51.652} {\bibfield
  {journal} {\bibinfo  {journal} {Phys. Rev.}\ }\textbf {\bibinfo {volume}
  {51}},\ \bibinfo {pages} {652} (\bibinfo {year} {1937})}\BibitemShut
  {NoStop}%
\bibitem [{\citenamefont {Xie}\ \emph {et~al.}(2017)\citenamefont {Xie},
  \citenamefont {Zhong}, \citenamefont {Batchelor},\ and\ \citenamefont
  {Lee}}]{Xie_2017}%
  \BibitemOpen
  \bibfield  {author} {\bibinfo {author} {\bibfnamefont {Q.}~\bibnamefont
  {Xie}}, \bibinfo {author} {\bibfnamefont {H.}~\bibnamefont {Zhong}}, \bibinfo
  {author} {\bibfnamefont {M.~T.}\ \bibnamefont {Batchelor}},\ and\ \bibinfo
  {author} {\bibfnamefont {C.}~\bibnamefont {Lee}},\ }\href
  {https://doi.org/10.1088/1751-8121/aa5a65} {\bibfield  {journal} {\bibinfo
  {journal} {J. Phys. A: Math. Theor.}\ }\textbf {\bibinfo {volume} {50}},\
  \bibinfo {pages} {113001} (\bibinfo {year} {2017})}\BibitemShut {NoStop}%
\bibitem [{\citenamefont {Toida}\ \emph {et~al.}(2013)\citenamefont {Toida},
  \citenamefont {Nakajima},\ and\ \citenamefont
  {Komiyama}}]{PhysRevLett.110.066802}%
  \BibitemOpen
  \bibfield  {author} {\bibinfo {author} {\bibfnamefont {H.}~\bibnamefont
  {Toida}}, \bibinfo {author} {\bibfnamefont {T.}~\bibnamefont {Nakajima}},\
  and\ \bibinfo {author} {\bibfnamefont {S.}~\bibnamefont {Komiyama}},\ }\href
  {https://doi.org/10.1103/PhysRevLett.110.066802} {\bibfield  {journal}
  {\bibinfo  {journal} {Phys. Rev. Lett.}\ }\textbf {\bibinfo {volume} {110}},\
  \bibinfo {pages} {066802} (\bibinfo {year} {2013})}\BibitemShut {NoStop}%
\bibitem [{\citenamefont {Yoshihara}\ \emph {et~al.}(2018)\citenamefont
  {Yoshihara}, \citenamefont {Fuse}, \citenamefont {Ao}, \citenamefont
  {Ashhab}, \citenamefont {Kakuyanagi}, \citenamefont {Saito}, \citenamefont
  {Aoki}, \citenamefont {Koshino},\ and\ \citenamefont
  {Semba}}]{Yoshihara_2018}%
  \BibitemOpen
  \bibfield  {author} {\bibinfo {author} {\bibfnamefont {F.}~\bibnamefont
  {Yoshihara}}, \bibinfo {author} {\bibfnamefont {T.}~\bibnamefont {Fuse}},
  \bibinfo {author} {\bibfnamefont {Z.}~\bibnamefont {Ao}}, \bibinfo {author}
  {\bibfnamefont {S.}~\bibnamefont {Ashhab}}, \bibinfo {author} {\bibfnamefont
  {K.}~\bibnamefont {Kakuyanagi}}, \bibinfo {author} {\bibfnamefont
  {S.}~\bibnamefont {Saito}}, \bibinfo {author} {\bibfnamefont
  {T.}~\bibnamefont {Aoki}}, \bibinfo {author} {\bibfnamefont {K.}~\bibnamefont
  {Koshino}},\ and\ \bibinfo {author} {\bibfnamefont {K.}~\bibnamefont
  {Semba}},\ }\href {https://doi.org/10.1103/PhysRevLett.120.183601} {\bibfield
   {journal} {\bibinfo  {journal} {Phys. Rev. Lett.}\ }\textbf {\bibinfo
  {volume} {120}},\ \bibinfo {pages} {183601} (\bibinfo {year}
  {2018})}\BibitemShut {NoStop}%
\bibitem [{\citenamefont {Brune}\ \emph {et~al.}(1996)\citenamefont {Brune},
  \citenamefont {Schmidt-Kaler}, \citenamefont {Maali}, \citenamefont {Dreyer},
  \citenamefont {Hagley}, \citenamefont {Raimond},\ and\ \citenamefont
  {Haroche}}]{PhysRevLett.76.1800}%
  \BibitemOpen
  \bibfield  {author} {\bibinfo {author} {\bibfnamefont {M.}~\bibnamefont
  {Brune}}, \bibinfo {author} {\bibfnamefont {F.}~\bibnamefont
  {Schmidt-Kaler}}, \bibinfo {author} {\bibfnamefont {A.}~\bibnamefont
  {Maali}}, \bibinfo {author} {\bibfnamefont {J.}~\bibnamefont {Dreyer}},
  \bibinfo {author} {\bibfnamefont {E.}~\bibnamefont {Hagley}}, \bibinfo
  {author} {\bibfnamefont {J.~M.}\ \bibnamefont {Raimond}},\ and\ \bibinfo
  {author} {\bibfnamefont {S.}~\bibnamefont {Haroche}},\ }\href
  {https://doi.org/10.1103/PhysRevLett.76.1800} {\bibfield  {journal} {\bibinfo
   {journal} {Phys. Rev. Lett.}\ }\textbf {\bibinfo {volume} {76}},\ \bibinfo
  {pages} {1800} (\bibinfo {year} {1996})}\BibitemShut {NoStop}%
\bibitem [{\citenamefont {Thompson}\ \emph {et~al.}(1992)\citenamefont
  {Thompson}, \citenamefont {Rempe},\ and\ \citenamefont
  {Kimble}}]{PhysRevLett.68.1132}%
  \BibitemOpen
  \bibfield  {author} {\bibinfo {author} {\bibfnamefont {R.~J.}\ \bibnamefont
  {Thompson}}, \bibinfo {author} {\bibfnamefont {G.}~\bibnamefont {Rempe}},\
  and\ \bibinfo {author} {\bibfnamefont {H.~J.}\ \bibnamefont {Kimble}},\
  }\href {https://doi.org/10.1103/PhysRevLett.68.1132} {\bibfield  {journal}
  {\bibinfo  {journal} {Phys. Rev. Lett.}\ }\textbf {\bibinfo {volume} {68}},\
  \bibinfo {pages} {1132} (\bibinfo {year} {1992})}\BibitemShut {NoStop}%
\bibitem [{\citenamefont {Trav{\v{e}}nec}\ and\ \citenamefont
  {{\v{S}}amaj}(2011)}]{Trav_nec_2011}%
  \BibitemOpen
  \bibfield  {author} {\bibinfo {author} {\bibfnamefont {I.}~\bibnamefont
  {Trav{\v{e}}nec}}\ and\ \bibinfo {author} {\bibfnamefont {L.}~\bibnamefont
  {{\v{S}}amaj}},\ }\href {https://doi.org/10.1016/j.physleta.2011.09.051}
  {\bibfield  {journal} {\bibinfo  {journal} {Phys. Lett. A}\ }\textbf
  {\bibinfo {volume} {375}},\ \bibinfo {pages} {4104} (\bibinfo {year}
  {2011})}\BibitemShut {NoStop}%
\bibitem [{\citenamefont {Crespi}\ \emph {et~al.}(2012)\citenamefont {Crespi},
  \citenamefont {Longhi},\ and\ \citenamefont {Osellame}}]{Crespi_2012}%
  \BibitemOpen
  \bibfield  {author} {\bibinfo {author} {\bibfnamefont {A.}~\bibnamefont
  {Crespi}}, \bibinfo {author} {\bibfnamefont {S.}~\bibnamefont {Longhi}},\
  and\ \bibinfo {author} {\bibfnamefont {R.}~\bibnamefont {Osellame}},\ }\href
  {https://doi.org/10.1103/PhysRevLett.108.163601} {\bibfield  {journal}
  {\bibinfo  {journal} {Phys. Rev. Lett.}\ }\textbf {\bibinfo {volume} {108}},\
  \bibinfo {pages} {163601} (\bibinfo {year} {2012})}\BibitemShut {NoStop}%
\bibitem [{\citenamefont {Kmetic}\ and\ \citenamefont
  {Meath}(1990)}]{PhysRevA.41.1556}%
  \BibitemOpen
  \bibfield  {author} {\bibinfo {author} {\bibfnamefont {M.~A.}\ \bibnamefont
  {Kmetic}}\ and\ \bibinfo {author} {\bibfnamefont {W.~J.}\ \bibnamefont
  {Meath}},\ }\href {https://doi.org/10.1103/PhysRevA.41.1556} {\bibfield
  {journal} {\bibinfo  {journal} {Phys. Rev. A}\ }\textbf {\bibinfo {volume}
  {41}},\ \bibinfo {pages} {1556} (\bibinfo {year} {1990})}\BibitemShut
  {NoStop}%
\bibitem [{\citenamefont {Brown}\ \emph {et~al.}(2002)\citenamefont {Brown},
  \citenamefont {Meath},\ and\ \citenamefont {Tran}}]{PhysRevA.65.063401}%
  \BibitemOpen
  \bibfield  {author} {\bibinfo {author} {\bibfnamefont {A.}~\bibnamefont
  {Brown}}, \bibinfo {author} {\bibfnamefont {W.~J.}\ \bibnamefont {Meath}},\
  and\ \bibinfo {author} {\bibfnamefont {P.}~\bibnamefont {Tran}},\ }\href
  {https://doi.org/10.1103/PhysRevA.65.063401} {\bibfield  {journal} {\bibinfo
  {journal} {Phys. Rev. A}\ }\textbf {\bibinfo {volume} {65}},\ \bibinfo
  {pages} {063401} (\bibinfo {year} {2002})}\BibitemShut {NoStop}%
\bibitem [{\citenamefont {Albert}(2012)}]{Albert_2012}%
  \BibitemOpen
  \bibfield  {author} {\bibinfo {author} {\bibfnamefont {V.~V.}\ \bibnamefont
  {Albert}},\ }\href {https://doi.org/10.1103/PhysRevLett.108.180401}
  {\bibfield  {journal} {\bibinfo  {journal} {Phys. Rev. Lett.}\ }\textbf
  {\bibinfo {volume} {108}},\ \bibinfo {pages} {180401} (\bibinfo {year}
  {2012})}\BibitemShut {NoStop}%
\bibitem [{\citenamefont {Leibfried}\ \emph {et~al.}(2003)\citenamefont
  {Leibfried}, \citenamefont {Blatt}, \citenamefont {Monroe},\ and\
  \citenamefont {Wineland}}]{RevModPhys.75.281}%
  \BibitemOpen
  \bibfield  {author} {\bibinfo {author} {\bibfnamefont {D.}~\bibnamefont
  {Leibfried}}, \bibinfo {author} {\bibfnamefont {R.}~\bibnamefont {Blatt}},
  \bibinfo {author} {\bibfnamefont {C.}~\bibnamefont {Monroe}},\ and\ \bibinfo
  {author} {\bibfnamefont {D.}~\bibnamefont {Wineland}},\ }\href
  {https://doi.org/10.1103/RevModPhys.75.281} {\bibfield  {journal} {\bibinfo
  {journal} {Rev. Mod. Phys.}\ }\textbf {\bibinfo {volume} {75}},\ \bibinfo
  {pages} {281} (\bibinfo {year} {2003})}\BibitemShut {NoStop}%
\bibitem [{\citenamefont {Pedernales}\ \emph {et~al.}(2015)\citenamefont
  {Pedernales}, \citenamefont {Lizuain}, \citenamefont {Felicetti},
  \citenamefont {Romero}, \citenamefont {Lamata},\ and\ \citenamefont
  {Solano}}]{Pedernales2015}%
  \BibitemOpen
  \bibfield  {author} {\bibinfo {author} {\bibfnamefont {J.~S.}\ \bibnamefont
  {Pedernales}}, \bibinfo {author} {\bibfnamefont {I.}~\bibnamefont {Lizuain}},
  \bibinfo {author} {\bibfnamefont {S.}~\bibnamefont {Felicetti}}, \bibinfo
  {author} {\bibfnamefont {G.}~\bibnamefont {Romero}}, \bibinfo {author}
  {\bibfnamefont {L.}~\bibnamefont {Lamata}},\ and\ \bibinfo {author}
  {\bibfnamefont {E.}~\bibnamefont {Solano}},\ }\href
  {https://doi.org/10.1038%2Fsrep15472} {\bibfield  {journal} {\bibinfo
  {journal} {Sci. Rep.}\ }\textbf {\bibinfo {volume} {5}} (\bibinfo {year}
  {2015})}\BibitemShut {NoStop}%
\bibitem [{\citenamefont {Cai}\ \emph {et~al.}(2021)\citenamefont {Cai},
  \citenamefont {Liu}, \citenamefont {Zhao}, \citenamefont {Wu}, \citenamefont
  {Mei}, \citenamefont {Jiang}, \citenamefont {He}, \citenamefont {Zhang},
  \citenamefont {Zhou},\ and\ \citenamefont {Duan}}]{Cai_2021}%
  \BibitemOpen
  \bibfield  {author} {\bibinfo {author} {\bibfnamefont {M.-L.}\ \bibnamefont
  {Cai}}, \bibinfo {author} {\bibfnamefont {Z.-D.}\ \bibnamefont {Liu}},
  \bibinfo {author} {\bibfnamefont {W.-D.}\ \bibnamefont {Zhao}}, \bibinfo
  {author} {\bibfnamefont {Y.-K.}\ \bibnamefont {Wu}}, \bibinfo {author}
  {\bibfnamefont {Q.-X.}\ \bibnamefont {Mei}}, \bibinfo {author} {\bibfnamefont
  {Y.}~\bibnamefont {Jiang}}, \bibinfo {author} {\bibfnamefont
  {L.}~\bibnamefont {He}}, \bibinfo {author} {\bibfnamefont {X.}~\bibnamefont
  {Zhang}}, \bibinfo {author} {\bibfnamefont {Z.-C.}\ \bibnamefont {Zhou}},\
  and\ \bibinfo {author} {\bibfnamefont {L.-M.}\ \bibnamefont {Duan}},\ }\href
  {https://doi.org/10.1038/s41467-021-21425-8} {\bibfield  {journal} {\bibinfo
  {journal} {Nat. Commun.}\ }\textbf {\bibinfo {volume} {12}} (\bibinfo {year}
  {2021})}\BibitemShut {NoStop}%
\bibitem [{\citenamefont {Lv}\ \emph {et~al.}(2018)\citenamefont {Lv},
  \citenamefont {An}, \citenamefont {Liu}, \citenamefont {Zhang}, \citenamefont
  {Pedernales}, \citenamefont {Lamata}, \citenamefont {Solano},\ and\
  \citenamefont {Kim}}]{Lv_2018}%
  \BibitemOpen
  \bibfield  {author} {\bibinfo {author} {\bibfnamefont {D.}~\bibnamefont
  {Lv}}, \bibinfo {author} {\bibfnamefont {S.}~\bibnamefont {An}}, \bibinfo
  {author} {\bibfnamefont {Z.}~\bibnamefont {Liu}}, \bibinfo {author}
  {\bibfnamefont {J.-N.}\ \bibnamefont {Zhang}}, \bibinfo {author}
  {\bibfnamefont {J.~S.}\ \bibnamefont {Pedernales}}, \bibinfo {author}
  {\bibfnamefont {L.}~\bibnamefont {Lamata}}, \bibinfo {author} {\bibfnamefont
  {E.}~\bibnamefont {Solano}},\ and\ \bibinfo {author} {\bibfnamefont
  {K.}~\bibnamefont {Kim}},\ }\href {https://doi.org/10.1103/PhysRevX.8.021027}
  {\bibfield  {journal} {\bibinfo  {journal} {Phys. Rev. X}\ }\textbf {\bibinfo
  {volume} {8}},\ \bibinfo {pages} {021027} (\bibinfo {year}
  {2018})}\BibitemShut {NoStop}%
\bibitem [{\citenamefont {Schmidt-Kaler}\ \emph {et~al.}(2003)\citenamefont
  {Schmidt-Kaler}, \citenamefont {Häffner}, \citenamefont {Riebe},
  \citenamefont {Gulde}, \citenamefont {Lancaster}, \citenamefont {Deuschle},
  \citenamefont {Becher}, \citenamefont {Roos}, \citenamefont {Eschner},\ and\
  \citenamefont {Blatt}}]{Schmidt_Kaler_2003}%
  \BibitemOpen
  \bibfield  {author} {\bibinfo {author} {\bibfnamefont {F.}~\bibnamefont
  {Schmidt-Kaler}}, \bibinfo {author} {\bibfnamefont {H.}~\bibnamefont
  {Häffner}}, \bibinfo {author} {\bibfnamefont {M.}~\bibnamefont {Riebe}},
  \bibinfo {author} {\bibfnamefont {S.}~\bibnamefont {Gulde}}, \bibinfo
  {author} {\bibfnamefont {G.~P.~T.}\ \bibnamefont {Lancaster}}, \bibinfo
  {author} {\bibfnamefont {T.}~\bibnamefont {Deuschle}}, \bibinfo {author}
  {\bibfnamefont {C.}~\bibnamefont {Becher}}, \bibinfo {author} {\bibfnamefont
  {C.~F.}\ \bibnamefont {Roos}}, \bibinfo {author} {\bibfnamefont
  {J.}~\bibnamefont {Eschner}},\ and\ \bibinfo {author} {\bibfnamefont
  {R.}~\bibnamefont {Blatt}},\ }\href {https://doi.org/10.1038/nature01494}
  {\bibfield  {journal} {\bibinfo  {journal} {Nature}\ }\textbf {\bibinfo
  {volume} {422}},\ \bibinfo {pages} {408} (\bibinfo {year}
  {2003})}\BibitemShut {NoStop}%
\bibitem [{\citenamefont {Moya-Cessa}(2016)}]{Moya_Cessa_2016}%
  \BibitemOpen
  \bibfield  {author} {\bibinfo {author} {\bibfnamefont {H.~M.}\ \bibnamefont
  {Moya-Cessa}},\ }\href {https://doi.org/10.1038/srep38961} {\bibfield
  {journal} {\bibinfo  {journal} {Sci. Rep.}\ }\textbf {\bibinfo {volume} {6}}
  (\bibinfo {year} {2016})}\BibitemShut {NoStop}%
\bibitem [{\citenamefont {Nielsen}\ and\ \citenamefont
  {Chuang}(2012)}]{Nielsen2012}%
  \BibitemOpen
  \bibfield  {author} {\bibinfo {author} {\bibfnamefont {M.~A.}\ \bibnamefont
  {Nielsen}}\ and\ \bibinfo {author} {\bibfnamefont {I.~L.}\ \bibnamefont
  {Chuang}},\ }\href {https://doi.org/10.1017/cbo9780511976667} {\emph
  {\bibinfo {title} {Quantum Computation and Quantum Information}}}\ (\bibinfo
  {publisher} {Cambridge University Press},\ \bibinfo {year}
  {2012})\BibitemShut {NoStop}%
\bibitem [{\citenamefont {Schuster}\ \emph {et~al.}(2005)\citenamefont
  {Schuster}, \citenamefont {Wallraff}, \citenamefont {Blais}, \citenamefont
  {Frunzio}, \citenamefont {Huang}, \citenamefont {Majer}, \citenamefont
  {Girvin},\ and\ \citenamefont {Schoelkopf}}]{PhysRevLett.94.123602}%
  \BibitemOpen
  \bibfield  {author} {\bibinfo {author} {\bibfnamefont {D.~I.}\ \bibnamefont
  {Schuster}}, \bibinfo {author} {\bibfnamefont {A.}~\bibnamefont {Wallraff}},
  \bibinfo {author} {\bibfnamefont {A.}~\bibnamefont {Blais}}, \bibinfo
  {author} {\bibfnamefont {L.}~\bibnamefont {Frunzio}}, \bibinfo {author}
  {\bibfnamefont {R.-S.}\ \bibnamefont {Huang}}, \bibinfo {author}
  {\bibfnamefont {J.}~\bibnamefont {Majer}}, \bibinfo {author} {\bibfnamefont
  {S.~M.}\ \bibnamefont {Girvin}},\ and\ \bibinfo {author} {\bibfnamefont
  {R.~J.}\ \bibnamefont {Schoelkopf}},\ }\href
  {https://doi.org/10.1103/PhysRevLett.94.123602} {\bibfield  {journal}
  {\bibinfo  {journal} {Phys. Rev. Lett.}\ }\textbf {\bibinfo {volume} {94}},\
  \bibinfo {pages} {123602} (\bibinfo {year} {2005})}\BibitemShut {NoStop}%
\bibitem [{\citenamefont {Lindström}\ \emph {et~al.}(2007)\citenamefont
  {Lindström}, \citenamefont {Webster}, \citenamefont {Healey}, \citenamefont
  {Colclough}, \citenamefont {Muirhead},\ and\ \citenamefont
  {Tzalenchuk}}]{Lindstr_m_2007}%
  \BibitemOpen
  \bibfield  {author} {\bibinfo {author} {\bibfnamefont {T.}~\bibnamefont
  {Lindström}}, \bibinfo {author} {\bibfnamefont {C.~H.}\ \bibnamefont
  {Webster}}, \bibinfo {author} {\bibfnamefont {J.~E.}\ \bibnamefont {Healey}},
  \bibinfo {author} {\bibfnamefont {M.~S.}\ \bibnamefont {Colclough}}, \bibinfo
  {author} {\bibfnamefont {C.~M.}\ \bibnamefont {Muirhead}},\ and\ \bibinfo
  {author} {\bibfnamefont {A.~Y.}\ \bibnamefont {Tzalenchuk}},\ }\href
  {https://doi.org/10.1088/0953-2048/20/8/016} {\bibfield  {journal} {\bibinfo
  {journal} {Supercond. Sci. Technol.}\ }\textbf {\bibinfo {volume} {20}},\
  \bibinfo {pages} {814} (\bibinfo {year} {2007})}\BibitemShut {NoStop}%
\bibitem [{\citenamefont {Aspelmeyer}\ \emph {et~al.}(2012)\citenamefont
  {Aspelmeyer}, \citenamefont {Meystre},\ and\ \citenamefont
  {Schwab}}]{Qoptomechanics}%
  \BibitemOpen
  \bibfield  {author} {\bibinfo {author} {\bibfnamefont {M.}~\bibnamefont
  {Aspelmeyer}}, \bibinfo {author} {\bibfnamefont {P.}~\bibnamefont
  {Meystre}},\ and\ \bibinfo {author} {\bibfnamefont {K.}~\bibnamefont
  {Schwab}},\ }\href {https://doi.org/10.1063/PT.3.1640} {\bibfield  {journal}
  {\bibinfo  {journal} {Phys. Today}\ }\textbf {\bibinfo {volume} {65}},\
  \bibinfo {pages} {29} (\bibinfo {year} {2012})}\BibitemShut {NoStop}%
\bibitem [{\citenamefont {{Corbitt}}\ and\ \citenamefont
  {{Mavalvala}}(2004)}]{2004JOptB...6S.675C}%
  \BibitemOpen
  \bibfield  {author} {\bibinfo {author} {\bibfnamefont {T.}~\bibnamefont
  {{Corbitt}}}\ and\ \bibinfo {author} {\bibfnamefont {N.}~\bibnamefont
  {{Mavalvala}}},\ }\href {https://doi.org/10.1088/1464-4266/6/8/008}
  {\bibfield  {journal} {\bibinfo  {journal} {J. Opt. B: Quantum Semiclass.
  Opt.}\ }\textbf {\bibinfo {volume} {6}},\ \bibinfo {pages} {S675} (\bibinfo
  {year} {2004})}\BibitemShut {NoStop}%
\bibitem [{\citenamefont {Abramovici}\ \emph {et~al.}(1992)\citenamefont
  {Abramovici}, \citenamefont {Althouse}, \citenamefont {Drever}, \citenamefont
  {G\"{u}rsel}, \citenamefont {Kawamura}, \citenamefont {Raab}, \citenamefont
  {Shoemaker}, \citenamefont {Sievers}, \citenamefont {Spero}, \citenamefont
  {Thorne}, \citenamefont {Vogt}, \citenamefont {Weiss}, \citenamefont
  {Whitcomb},\ and\ \citenamefont {Zucker}}]{Abramovici1992}%
  \BibitemOpen
  \bibfield  {author} {\bibinfo {author} {\bibfnamefont {A.}~\bibnamefont
  {Abramovici}}, \bibinfo {author} {\bibfnamefont {W.~E.}\ \bibnamefont
  {Althouse}}, \bibinfo {author} {\bibfnamefont {R.~W.~P.}\ \bibnamefont
  {Drever}}, \bibinfo {author} {\bibfnamefont {Y.}~\bibnamefont {G\"{u}rsel}},
  \bibinfo {author} {\bibfnamefont {S.}~\bibnamefont {Kawamura}}, \bibinfo
  {author} {\bibfnamefont {F.~J.}\ \bibnamefont {Raab}}, \bibinfo {author}
  {\bibfnamefont {D.}~\bibnamefont {Shoemaker}}, \bibinfo {author}
  {\bibfnamefont {L.}~\bibnamefont {Sievers}}, \bibinfo {author} {\bibfnamefont
  {R.~E.}\ \bibnamefont {Spero}}, \bibinfo {author} {\bibfnamefont {K.~S.}\
  \bibnamefont {Thorne}}, \bibinfo {author} {\bibfnamefont {R.~E.}\
  \bibnamefont {Vogt}}, \bibinfo {author} {\bibfnamefont {R.}~\bibnamefont
  {Weiss}}, \bibinfo {author} {\bibfnamefont {S.~E.}\ \bibnamefont
  {Whitcomb}},\ and\ \bibinfo {author} {\bibfnamefont {M.~E.}\ \bibnamefont
  {Zucker}},\ }\href {https://doi.org/10.1126/science.256.5055.325} {\bibfield
  {journal} {\bibinfo  {journal} {Science}\ }\textbf {\bibinfo {volume}
  {256}},\ \bibinfo {pages} {325} (\bibinfo {year} {1992})}\BibitemShut
  {NoStop}%
\bibitem [{\citenamefont {Ma}\ \emph {et~al.}(2021)\citenamefont {Ma},
  \citenamefont {Viennot}, \citenamefont {Kotler}, \citenamefont {Teufel},\
  and\ \citenamefont {Lehnert}}]{Non-classical}%
  \BibitemOpen
  \bibfield  {author} {\bibinfo {author} {\bibfnamefont {X.}~\bibnamefont
  {Ma}}, \bibinfo {author} {\bibfnamefont {J.}~\bibnamefont {Viennot}},
  \bibinfo {author} {\bibfnamefont {S.}~\bibnamefont {Kotler}}, \bibinfo
  {author} {\bibfnamefont {J.}~\bibnamefont {Teufel}},\ and\ \bibinfo {author}
  {\bibfnamefont {K.}~\bibnamefont {Lehnert}},\ }\href
  {https://doi.org/10.1038/s41567-020-01102-1} {\bibfield  {journal} {\bibinfo
  {journal} {Nat. Phys.}\ }\textbf {\bibinfo {volume} {17}},\ \bibinfo {pages}
  {1} (\bibinfo {year} {2021})}\BibitemShut {NoStop}%
\bibitem [{\citenamefont {Restrepo}\ \emph {et~al.}(2014)\citenamefont
  {Restrepo}, \citenamefont {Ciuti},\ and\ \citenamefont
  {Favero}}]{PhysRevLett.112.013601}%
  \BibitemOpen
  \bibfield  {author} {\bibinfo {author} {\bibfnamefont {J.}~\bibnamefont
  {Restrepo}}, \bibinfo {author} {\bibfnamefont {C.}~\bibnamefont {Ciuti}},\
  and\ \bibinfo {author} {\bibfnamefont {I.}~\bibnamefont {Favero}},\ }\href
  {https://doi.org/10.1103/PhysRevLett.112.013601} {\bibfield  {journal}
  {\bibinfo  {journal} {Phys. Rev. Lett.}\ }\textbf {\bibinfo {volume} {112}},\
  \bibinfo {pages} {013601} (\bibinfo {year} {2014})}\BibitemShut {NoStop}%
\bibitem [{\citenamefont {Restrepo}\ \emph {et~al.}(2017)\citenamefont
  {Restrepo}, \citenamefont {Favero},\ and\ \citenamefont
  {Ciuti}}]{PhysRevA.95.023832}%
  \BibitemOpen
  \bibfield  {author} {\bibinfo {author} {\bibfnamefont {J.}~\bibnamefont
  {Restrepo}}, \bibinfo {author} {\bibfnamefont {I.}~\bibnamefont {Favero}},\
  and\ \bibinfo {author} {\bibfnamefont {C.}~\bibnamefont {Ciuti}},\ }\href
  {https://doi.org/10.1103/PhysRevA.95.023832} {\bibfield  {journal} {\bibinfo
  {journal} {Phys. Rev. A}\ }\textbf {\bibinfo {volume} {95}},\ \bibinfo
  {pages} {023832} (\bibinfo {year} {2017})}\BibitemShut {NoStop}%
\bibitem [{\citenamefont {Shore}\ and\ \citenamefont
  {Knight}(1993)}]{Shore_1993}%
  \BibitemOpen
  \bibfield  {author} {\bibinfo {author} {\bibfnamefont {B.~W.}\ \bibnamefont
  {Shore}}\ and\ \bibinfo {author} {\bibfnamefont {P.~L.}\ \bibnamefont
  {Knight}},\ }\href {https://doi.org/10.1080/09500349314551321} {\bibfield
  {journal} {\bibinfo  {journal} {J. Mod. Opt.}\ }\textbf {\bibinfo {volume}
  {40}},\ \bibinfo {pages} {1195} (\bibinfo {year} {1993})}\BibitemShut
  {NoStop}%
\bibitem [{\citenamefont {{Bina}}(2012)}]{2012EPJST.203..163B}%
  \BibitemOpen
  \bibfield  {author} {\bibinfo {author} {\bibfnamefont {M.}~\bibnamefont
  {{Bina}}},\ }\href {https://doi.org/10.1140/epjst/e2012-01541-3} {\bibfield
  {journal} {\bibinfo  {journal} {Eur. Phys. J. Spec. Top.}\ }\textbf {\bibinfo
  {volume} {203}},\ \bibinfo {pages} {163} (\bibinfo {year}
  {2012})}\BibitemShut {NoStop}%
\bibitem [{\citenamefont {Braak}(2019)}]{Braak_2019}%
  \BibitemOpen
  \bibfield  {author} {\bibinfo {author} {\bibfnamefont {D.}~\bibnamefont
  {Braak}},\ }\href {https://doi.org/10.3390/sym11101259} {\bibfield  {journal}
  {\bibinfo  {journal} {Symmetry}\ }\textbf {\bibinfo {volume} {11}},\ \bibinfo
  {pages} {1259} (\bibinfo {year} {2019})}\BibitemShut {NoStop}%
\bibitem [{\citenamefont {Cummings}(2013)}]{JC_Journal}%
  \BibitemOpen
  \bibfield  {author} {\bibinfo {author} {\bibfnamefont {F.~W.}\ \bibnamefont
  {Cummings}},\ }\href {https://doi.org/10.1088/0953-4075/46/22/220202}
  {\bibfield  {journal} {\bibinfo  {journal} {J. Phys. B: At. Mol. Opt. Phys.}\
  }\textbf {\bibinfo {volume} {46}},\ \bibinfo {pages} {220202} (\bibinfo
  {year} {2013})}\BibitemShut {NoStop}%
\bibitem [{\citenamefont {Forn-D\'{\i}az}\ \emph {et~al.}(2019)\citenamefont
  {Forn-D\'{\i}az}, \citenamefont {Lamata}, \citenamefont {Rico}, \citenamefont
  {Kono},\ and\ \citenamefont {Solano}}]{RevModPhys.91.025005}%
  \BibitemOpen
  \bibfield  {author} {\bibinfo {author} {\bibfnamefont {P.}~\bibnamefont
  {Forn-D\'{\i}az}}, \bibinfo {author} {\bibfnamefont {L.}~\bibnamefont
  {Lamata}}, \bibinfo {author} {\bibfnamefont {E.}~\bibnamefont {Rico}},
  \bibinfo {author} {\bibfnamefont {J.}~\bibnamefont {Kono}},\ and\ \bibinfo
  {author} {\bibfnamefont {E.}~\bibnamefont {Solano}},\ }\href
  {https://doi.org/10.1103/RevModPhys.91.025005} {\bibfield  {journal}
  {\bibinfo  {journal} {Rev. Mod. Phys.}\ }\textbf {\bibinfo {volume} {91}},\
  \bibinfo {pages} {025005} (\bibinfo {year} {2019})}\BibitemShut {NoStop}%
\bibitem [{\citenamefont {Kockum}\ \emph {et~al.}(2019)\citenamefont {Kockum},
  \citenamefont {Miranowicz}, \citenamefont {Liberato}, \citenamefont
  {Savasta},\ and\ \citenamefont {Nori}}]{Frisk_Kockum_2019}%
  \BibitemOpen
  \bibfield  {author} {\bibinfo {author} {\bibfnamefont {A.~F.}\ \bibnamefont
  {Kockum}}, \bibinfo {author} {\bibfnamefont {A.}~\bibnamefont {Miranowicz}},
  \bibinfo {author} {\bibfnamefont {S.~D.}\ \bibnamefont {Liberato}}, \bibinfo
  {author} {\bibfnamefont {S.}~\bibnamefont {Savasta}},\ and\ \bibinfo {author}
  {\bibfnamefont {F.}~\bibnamefont {Nori}},\ }\href
  {https://doi.org/10.1038/s42254-018-0006-2} {\bibfield  {journal} {\bibinfo
  {journal} {Nat. Rev. Phys.}\ }\textbf {\bibinfo {volume} {1}},\ \bibinfo
  {pages} {19} (\bibinfo {year} {2019})}\BibitemShut {NoStop}%
\bibitem [{\citenamefont {Irish}\ \emph {et~al.}(2005)\citenamefont {Irish},
  \citenamefont {Gea-Banacloche}, \citenamefont {Martin},\ and\ \citenamefont
  {Schwab}}]{PhysRevB.72.195410}%
  \BibitemOpen
  \bibfield  {author} {\bibinfo {author} {\bibfnamefont {E.~K.}\ \bibnamefont
  {Irish}}, \bibinfo {author} {\bibfnamefont {J.}~\bibnamefont
  {Gea-Banacloche}}, \bibinfo {author} {\bibfnamefont {I.}~\bibnamefont
  {Martin}},\ and\ \bibinfo {author} {\bibfnamefont {K.~C.}\ \bibnamefont
  {Schwab}},\ }\href {https://doi.org/10.1103/PhysRevB.72.195410} {\bibfield
  {journal} {\bibinfo  {journal} {Phys. Rev. B}\ }\textbf {\bibinfo {volume}
  {72}},\ \bibinfo {pages} {195410} (\bibinfo {year} {2005})}\BibitemShut
  {NoStop}%
\bibitem [{\citenamefont {Irish}(2007)}]{PhysRevLett.99.173601}%
  \BibitemOpen
  \bibfield  {author} {\bibinfo {author} {\bibfnamefont {E.~K.}\ \bibnamefont
  {Irish}},\ }\href {https://doi.org/10.1103/PhysRevLett.99.173601} {\bibfield
  {journal} {\bibinfo  {journal} {Phys. Rev. Lett.}\ }\textbf {\bibinfo
  {volume} {99}},\ \bibinfo {pages} {173601} (\bibinfo {year}
  {2007})}\BibitemShut {NoStop}%
\bibitem [{\citenamefont {Albert}\ \emph {et~al.}(2011)\citenamefont {Albert},
  \citenamefont {Scholes},\ and\ \citenamefont {Brumer}}]{PhysRevA.84.042110}%
  \BibitemOpen
  \bibfield  {author} {\bibinfo {author} {\bibfnamefont {V.~V.}\ \bibnamefont
  {Albert}}, \bibinfo {author} {\bibfnamefont {G.~D.}\ \bibnamefont
  {Scholes}},\ and\ \bibinfo {author} {\bibfnamefont {P.}~\bibnamefont
  {Brumer}},\ }\href {https://doi.org/10.1103/PhysRevA.84.042110} {\bibfield
  {journal} {\bibinfo  {journal} {Phys. Rev. A}\ }\textbf {\bibinfo {volume}
  {84}},\ \bibinfo {pages} {042110} (\bibinfo {year} {2011})}\BibitemShut
  {NoStop}%
\bibitem [{\citenamefont {Sandu}(2015)}]{disp1}%
  \BibitemOpen
  \bibfield  {author} {\bibinfo {author} {\bibfnamefont {T.}~\bibnamefont
  {Sandu}},\ }\href {https://rjp.nipne.ro/2015_60_5-6/RomJPhys.60.p711.pdf}
  {\bibfield  {journal} {\bibinfo  {journal} {Rom. J. Phys.}\ }\textbf
  {\bibinfo {volume} {60}},\ \bibinfo {pages} {711} (\bibinfo {year}
  {2015})}\BibitemShut {NoStop}%
\bibitem [{\citenamefont {Casanova}\ \emph {et~al.}(2010)\citenamefont
  {Casanova}, \citenamefont {Romero}, \citenamefont {Lizuain}, \citenamefont
  {Garc\'{\i}a-Ripoll},\ and\ \citenamefont {Solano}}]{PhysRevLett.105.263603}%
  \BibitemOpen
  \bibfield  {author} {\bibinfo {author} {\bibfnamefont {J.}~\bibnamefont
  {Casanova}}, \bibinfo {author} {\bibfnamefont {G.}~\bibnamefont {Romero}},
  \bibinfo {author} {\bibfnamefont {I.}~\bibnamefont {Lizuain}}, \bibinfo
  {author} {\bibfnamefont {J.~J.}\ \bibnamefont {Garc\'{\i}a-Ripoll}},\ and\
  \bibinfo {author} {\bibfnamefont {E.}~\bibnamefont {Solano}},\ }\href
  {https://doi.org/10.1103/PhysRevLett.105.263603} {\bibfield  {journal}
  {\bibinfo  {journal} {Phys. Rev. Lett.}\ }\textbf {\bibinfo {volume} {105}},\
  \bibinfo {pages} {263603} (\bibinfo {year} {2010})}\BibitemShut {NoStop}%
\bibitem [{\citenamefont {Braak}(2011)}]{PhysRevLett.107.100401}%
  \BibitemOpen
  \bibfield  {author} {\bibinfo {author} {\bibfnamefont {D.}~\bibnamefont
  {Braak}},\ }\href {https://doi.org/10.1103/PhysRevLett.107.100401} {\bibfield
   {journal} {\bibinfo  {journal} {Phys. Rev. Lett.}\ }\textbf {\bibinfo
  {volume} {107}},\ \bibinfo {pages} {100401} (\bibinfo {year}
  {2011})}\BibitemShut {NoStop}%
\bibitem [{\citenamefont {Solano}(2011)}]{solano2011dialogue}%
  \BibitemOpen
  \bibfield  {author} {\bibinfo {author} {\bibfnamefont {E.}~\bibnamefont
  {Solano}},\ }\href {https://doi.org/10.1103/Physics.4.68} {\bibfield
  {journal} {\bibinfo  {journal} {Physics}\ }\textbf {\bibinfo {volume} {4}},\
  \bibinfo {pages} {68} (\bibinfo {year} {2011})}\BibitemShut {NoStop}%
\bibitem [{\citenamefont {Braginsky}(1967)}]{braginsky1967classical}%
  \BibitemOpen
  \bibfield  {author} {\bibinfo {author} {\bibfnamefont {V.}~\bibnamefont
  {Braginsky}},\ }\href {http://jetp.ras.ru/cgi-bin/dn/e_026_04_0831.pdf}
  {\bibfield  {journal} {\bibinfo  {journal} {Zh. Eksp. Teor. Fiz}\ }\textbf
  {\bibinfo {volume} {53}},\ \bibinfo {pages} {1434} (\bibinfo {year}
  {1967})}\BibitemShut {NoStop}%
\bibitem [{\citenamefont {Jaynes}\ and\ \citenamefont
  {Cummings}(1963)}]{1443594}%
  \BibitemOpen
  \bibfield  {author} {\bibinfo {author} {\bibfnamefont {E.}~\bibnamefont
  {Jaynes}}\ and\ \bibinfo {author} {\bibfnamefont {F.}~\bibnamefont
  {Cummings}},\ }\href {https://doi.org/10.1109/PROC.1963.1664} {\bibfield
  {journal} {\bibinfo  {journal} {Proceedings of the IEEE}\ }\textbf {\bibinfo
  {volume} {51}},\ \bibinfo {pages} {89} (\bibinfo {year} {1963})}\BibitemShut
  {NoStop}%
\bibitem [{\citenamefont {Benenti}\ \emph {et~al.}(2007)\citenamefont
  {Benenti}, \citenamefont {Casati},\ and\ \citenamefont {Strini}}]{Giuliano}%
  \BibitemOpen
  \bibfield  {author} {\bibinfo {author} {\bibfnamefont {G.}~\bibnamefont
  {Benenti}}, \bibinfo {author} {\bibfnamefont {G.}~\bibnamefont {Casati}},\
  and\ \bibinfo {author} {\bibfnamefont {G.}~\bibnamefont {Strini}},\ }\href
  {https://doi.org/10.1142/5838} {\emph {\bibinfo {title} {{P}rinciples of
  {Q}uantum {C}omputation and {I}nformation}}}\ (\bibinfo  {publisher} {{WORLD}
  {SCIENTIFIC}},\ \bibinfo {year} {2007})\BibitemShut {NoStop}%
\bibitem [{\citenamefont {Liao}\ \emph {et~al.}(2018)\citenamefont {Liao},
  \citenamefont {Ye}, \citenamefont {Jin}, \citenamefont {Zhou},\ and\
  \citenamefont {Nie}}]{Liao2018}%
  \BibitemOpen
  \bibfield  {author} {\bibinfo {author} {\bibfnamefont {Q.}~\bibnamefont
  {Liao}}, \bibinfo {author} {\bibfnamefont {Y.}~\bibnamefont {Ye}}, \bibinfo
  {author} {\bibfnamefont {P.}~\bibnamefont {Jin}}, \bibinfo {author}
  {\bibfnamefont {N.}~\bibnamefont {Zhou}},\ and\ \bibinfo {author}
  {\bibfnamefont {W.}~\bibnamefont {Nie}},\ }\href
  {https://doi.org/10.1007/s10773-017-3661-7} {\bibfield  {journal} {\bibinfo
  {journal} {Int. J. Theor. Phys.}\ }\textbf {\bibinfo {volume} {57}},\
  \bibinfo {pages} {1319} (\bibinfo {year} {2018})}\BibitemShut {NoStop}%
\bibitem [{\citenamefont {Haroche}\ and\ \citenamefont
  {Raimond}(2006)}]{haroche_book}%
  \BibitemOpen
  \bibfield  {author} {\bibinfo {author} {\bibfnamefont {S.}~\bibnamefont
  {Haroche}}\ and\ \bibinfo {author} {\bibfnamefont {J.-M.}\ \bibnamefont
  {Raimond}},\ }\href
  {https://doi.org/10.1093/acprof:oso/9780198509141.001.0001} {\emph {\bibinfo
  {title} {{Exploring the {Q}uantum: {A}toms, {C}avities, and {P}hotons}}}}\
  (\bibinfo  {publisher} {Oxford University Press},\ \bibinfo {year}
  {2006})\BibitemShut {NoStop}%
\end{thebibliography}%

\end{document}